%% file: OHIO_solar.tex
\documentclass[twocolumn]{aastex631}

\usepackage{amsmath}
\usepackage{wrapfig}
\usepackage{physics}
\usepackage{tikz}
\let\tablewidth\relax
\usepackage{tabularray}
\usepackage{mathdots}
\usepackage{yhmath}
\usepackage{cancel}
\usepackage{color}
\usepackage{xcolor}
\let\tablenum\relax
\usepackage{siunitx}
\usepackage{array}
\usepackage{multirow}
\usepackage{amssymb}
\usepackage{gensymb}
\usepackage{tabularx}
\usepackage{extarrows}
\usepackage{booktabs}
\usepackage{appendix}
\usetikzlibrary{fadings}
\usetikzlibrary{patterns}
\usetikzlibrary{shadows.blur}
\usetikzlibrary{shapes}
\usepackage{url}
\usepackage{wrapfig}

\begin{document}

\title{Constraining solar emission radius at 42 MHz during the 2024 total solar eclipse using a student-commissioned radio telescope }

\author[0000-0002-0883-0688]{Olivia R. Young}
\affiliation{School of Physics and Astronomy, Rochester Institute of Technology, Rochester, NY 14623, USA}
\affiliation{National Radio Astronomy Observatory, 520 Edgemont Road, Charlottesville, VA 22903, USA}
\affiliation{Center for Detectors, Rochester Institute of Technology, Rochester, NY 14623, USA}
\affiliation{Laboratory for Multiwavelength Astrophysics, Rochester Institute of Technology, Rochester, NY 14623, USA}

\author[0000-0001-8885-6388]{Timothy E. Dolch}
\affiliation{Department of Physics, Hillsdale College, 33 E. College Street, Hillsdale, MI 49242, USA}
\affiliation{Eureka Scientific, 2452 Delmer Street, Suite 100, Oakland, CA 94602-3017, USA }

\author[0000-0002-2123-0441]{Joseph F. Helmboldt}
\affiliation{U.S. Naval Research Laboratory, 4555 Overlook Ave, SW, Washington, DC 20375, USA
}

\author{Christopher Mentrek}
\affiliation{Observatory Park, Geauga Parks District, Chardon, OH 44064, USA
}

\author[0000-0002-2216-0465]{Louis P. Dartez}
\affiliation{LIGO Hanford Observatory, Richland, WA 99352, USA
}

\author[0000-0003-0721-651X]{Michael T. Lam}
\affiliation{SETI Institute, 339 N Bernardo Ave Suite 200, Mountain View, CA 94043, USA}
\affiliation{School of Physics and Astronomy, Rochester Institute of Technology, Rochester, NY 14623, USA}
\affiliation{Laboratory for Multiwavelength Astrophysics, Rochester Institute of Technology, Rochester, NY
14623, USA}

\author[0000-0002-5176-2924]{Sophia V. Sosa Fiscella}
\affiliation{School of Physics and Astronomy, Rochester Institute of Technology, Rochester, NY 14623, USA}
\affiliation{Laboratory for Multiwavelength Astrophysics, Rochester Institute of Technology, Rochester, NY 14623, USA}

\author{Evan Bretl}
\affiliation{Society of Amateur Radio Astronomers, Charlottesville, VA, USA}

\author[0009-0002-1763-9962]{Colin Joyce}
\affiliation{Department of Physics, Hillsdale College, 33 E. College Street, Hillsdale, MI 49242, USA}

\author[0009-0008-8163-6495]{Johannes Loock}
\affiliation{Department of Physics, Hillsdale College, 33 E. College Street, Hillsdale, MI 49242, USA}

\author[0009-0008-8755-5266]{Grace Meyer}
\affiliation{Department of Physics, Hillsdale College, 33 E. College Street, Hillsdale, MI 49242, USA}

\author[0009-0000-4697-4319]{Annabel Peltzer}
\affiliation{Department of Physics, Hillsdale College, 33 E. College Street, Hillsdale, MI 49242, USA}

\author[0000-0001-9132-2942]{Joseph Petullo}
\affiliation{Department of Physics, Hillsdale College, 33 E. College Street, Hillsdale, MI 49242, USA}

\author[0000-0002-2410-6227]{Parker Reed}
\affiliation{Department of Physics, Hillsdale College, 33 E. College Street, Hillsdale, MI 49242, USA}

\author{Emerson Sigtryggsson}
\affiliation{Department of Physics, Hillsdale College, 33 E. College Street, Hillsdale, MI 49242, USA}

\author[0009-0009-2609-2445]{Benjamin Bassett}
\affiliation{Department of Physics, Hillsdale College, 33 E. College Street, Hillsdale, MI 49242, USA}

\author[0009-0004-2212-7771]{Andrew B. Hawken}
\affiliation{Department of Physics, Hillsdale College, 33 E. College Street, Hillsdale, MI 49242, USA}

\author[0009-0008-8151-2978]{Alejandro Z. Heredia}
\affiliation{Department of Physics, Hillsdale College, 33 E. College Street, Hillsdale, MI 49242, USA}

\author[0000-0001-5061-2281]{Paige Lettow}
\affiliation{Department of Physics, Hillsdale College, 33 E. College Street, Hillsdale, MI 49242, USA}

\author[0000-0002-6540-1863]{Whit Lewis}
\affiliation{Department of Physics, Hillsdale College, 33 E. College Street, Hillsdale, MI 49242, USA}

\author[0009-0000-0411-6068]{Mikayla Manna}
\affiliation{Department of Physics, Hillsdale College, 33 E. College Street, Hillsdale, MI 49242, USA}

\author[0009-0005-5249-8562]{Nicholas Mirochnikoff}
\affiliation{Department of Physics, Hillsdale College, 33 E. College Street, Hillsdale, MI 49242, USA}

\author[0000-0001-8253-1451]{Michael Zemcov}
\affiliation{School of Physics and Astronomy, Rochester Institute of Technology, Rochester, NY 14623, USA} 
\affiliation{Jet Propulsion Laboratory, 4800 Oak Grove Dr., Pasadena, CA 91109, USA}



\begin{abstract}

Low-frequency solar radio emission is sourced in the solar corona, with sub-100 MHz radio emission largely originating from the $\sim$10$^{5}$\,$\mathrm{K}$ plasma around 2 optical radii.  However, the region of emission has yet to be constrained at 35--45\,MHz due to both instrumentation limitations and the rarity of astronomical events, such as total solar eclipses, which allow for direct observational approaches. In this work, we present the results from a student-led project to commission a low-frequency radio telescope array situated in the path of totality of the 2024 total solar eclipse in an effort to probe the middle corona. The Deployable Low-Band Ionosphere and Transient Experiment (DLITE) is a low-frequency radio array comprised of four dipole antennas, optimized to observe at 35--45\,MHz, and capable of resolving the brightest radio sources in the sky. We constructed a DLITE station in Observatory Park, a dark sky park in Montville, Ohio. Results of observations during the total solar eclipse demonstrate that DLITE stations can be quickly deployed for observations and provide constraints on the radius of solar emission at our center observing frequency of 42\,MHz. In this work, we outline the construction of DLITE Ohio and the solar observation results from the total solar eclipse that transversed North America in April 2024.

\end{abstract}


\section{Introduction}

Solar radio radiation from the quiet Sun is dominated by the thermal bremsstrahlung emission generated in the middle and upper corona, resulting in an extended and non-spherical region of radio emission due to the high plasma temperatures \citep{Vocks_2018}. However, this is not the case for emission at the lowest radio frequencies, particularly in the $1{-}100$\,MHz range. Low-frequency radio emission in the middle corona, defined as the region from $1.5 {-} 6$\,$R_\odot$, is dominated by plasma emission \citep{West_2023, Gary2005}. In the $30 {-} 45$\,MHz range, radio solar emission is expected to originate from about 1 optical solar radius above the photosphere of the Sun, as evidenced by low-frequency radio emission ($30{-}100$\,MHz) accompanying coronal mass ejections \citep{2014ApJ, Mugundhan_2016, West_2023, Gary2005, white2024solarradioburstsspace}. During times in which the Sun is active, solar radio intensity fluctuates drastically on both short- and long-time scales due to a variety of radio bursts associated with active regions, coronal mass ejections, coronal loops, and solar flares \citep{Lofar_bursts, flares, Zhang_2024, 1963ARA&A...1..291W}.

Due to both instrumentation constraints, namely the need for baselines on the order of 1\,km or greater, and the heightened effects of scattering at low radio frequencies, the current body of work on solar low-frequency radio emission is largely restricted to observations of the quiet Sun and its active regions \citep{LoFar, 1994ApJ...426..774B}. Therefore, observations at very low frequencies below the FM radio band ($30{-}45$\,MHz), especially during solar maximum, stand to provide insight into these poorly understood emission regions of the solar atmosphere.   

Observations by the LOw Frequency ARray (LOFAR), a telescope comprised of thousands of antennas located throughout Europe with maximum baselines of $\sim$2000\,$\mathrm{km}$\, \citep{LOFAR_com}, during a partial solar eclipse in March 2015 at 120--180\,MHz showed the efficacy of using rare events such as eclipses as tools to work past the limitations inherent to the lower radio frequencies. These observations were taken during a solar minimum, leading to observations of a quiet solar corona at 80$\%$ totality \citep{LoFar}. An innovative lunar de-occultation technique (see \cite{10.1093/mnras/stv746} and \cite{LoFar} for more details) was used to achieve $\sim$0.6$'$ resolution, resolving structures as small as a few arcminutes in the solar corona. When compared to data on which the de-occultation technique had not been employed, the authors found that sub-arcminute structures were not observed, agreeing with previous studies that claimed that such structures are unobservable regardless of angular resolution due to scattering effects at low frequencies \citep{1994ApJ...426..774B}. Thus, attempts at increasing resolution at these frequencies past the arc-minute resolution achieved by LOFAR will not result in increased sensitivity to solar structures. 

This resolution ceiling at low radio frequencies opens the door for continued contributions from telescopes with smaller baselines and resolutions larger than an arcminute, as evidenced by the solar observation efforts by the Long Wavelength Array \citep[LWA;][]{2012PASP..124.1090H}. Observations with the LWA station at ORVO achieve spatial resolutions in the range of $5'$ - $25'$ across its band ( $\sim$6$'$ at 88 MHz) and is used for solar emission mapping \citep{2021AGUFMSH52A..03G}. Thus, a large range of resolutions are sufficient to resolve the sun at low radio frequencies and provide insight into low-frequency solar emission \citep{West_2023}.

Due to the large numbers of antennas and huge baselines needed to achieve a similar resolution as LOFAR or the LWA, arrays of this type are not easily deployable to take advantage of location-dependent events such as total or partial solar eclipses. Thus, up until now, there have been no known low-frequency radio observations of the solar corona during a solar maximum from within the path of totality of an eclipse.  

The Deployable Low-Band Ionosphere and Transient Experiment (DLITE) is a low-frequency deployable interferometer designed for operation between 35 and 45\,MHz and was designed with the option of quick installation to observe rare phenomena, with the primary focus on observing changes to the ionosphere \citep{DLITE}. DLITE is also able to probe low-frequency radio emissions from the Sun and is especially proficient at studying solar radio bursts (SRBs) \citep{DLITE_Solar}. The DLITE system has shorter baselines by several orders of magnitude compared to LOFAR, resulting in a resolution on the order of 1$\degree$ for visibility images of the sky at 35--45\,MHz. The extent of solar radiation on the sky at these frequencies is considerably larger than in optical ranges, resulting in the footprint of the low-frequency radio Sun appearing between 1.2 and 10 optical solar radii between 100\,MHz and 1\,MHz, respectively \citep{West_2023, Gary2005}. Therefore, an instrument such as DLITE that can be quickly and affordably installed in the path of totality of a solar eclipse would enable confirmation of the radius of emission at these frequencies.

In section \ref{sec:DLITE_system}, we describe the DLITE system; in section \ref{sec: com}, we detail the student-led commissioning of a DLITE array in Observatory Park in Montville, Ohio; in section \ref{sec: obs}, we discuss results from the 2024 total solar eclipse as seen with the DLITE array in Ohio; and in section \ref{sec: outreach} we discuss the development of DLITE TV as an outreach tool to live stream our observation of the eclipse, as well as the development of an open-source data processing notebook.

\section{The DLITE Ohio System Implementation}
\label{sec:DLITE_system}

The DLITE system, as originally described in \cite{DLITE}, is an interferometric radio telescope comprised of four LWA-style dipole antennas and built primarily from commercially available, off-the-shelf parts. This makes DLITE well-suited for use in student-led projects at the university level and facilitates rapid deployment in preparation for opportunistic observations of rare phenomena such as the 2024 total solar eclipse in North America. Here we briefly review the system architecture and outline characteristics unique to the system deployed in Ohio.

DLITE is optimized to operate in the 35--45\,MHz band and to probe the structure of the ionosphere using bright radio sources such as the ``A-Team'': Cassiopeia A, Cygnus A, Taurus A, Virgo A, Hercules A, and Hydra A. DLITE tracks these bright sources to both measure ionospheric scintillations due to km-scale density irregularities and total electron count (TEC) gradients that are typically driven by medium-scale ($\sim$50--300\,km) disturbances \citep{https://doi.org/10.1029/2021RS007396}. DLITE is also particularly well-suited for constant solar emission monitoring, both because of its ability to detect SRBs and low data rates \citep{DLITE_Solar, DLITE}.  

 \begin{figure}[h!]
    \centering
         \includegraphics[width = 1.0\linewidth]{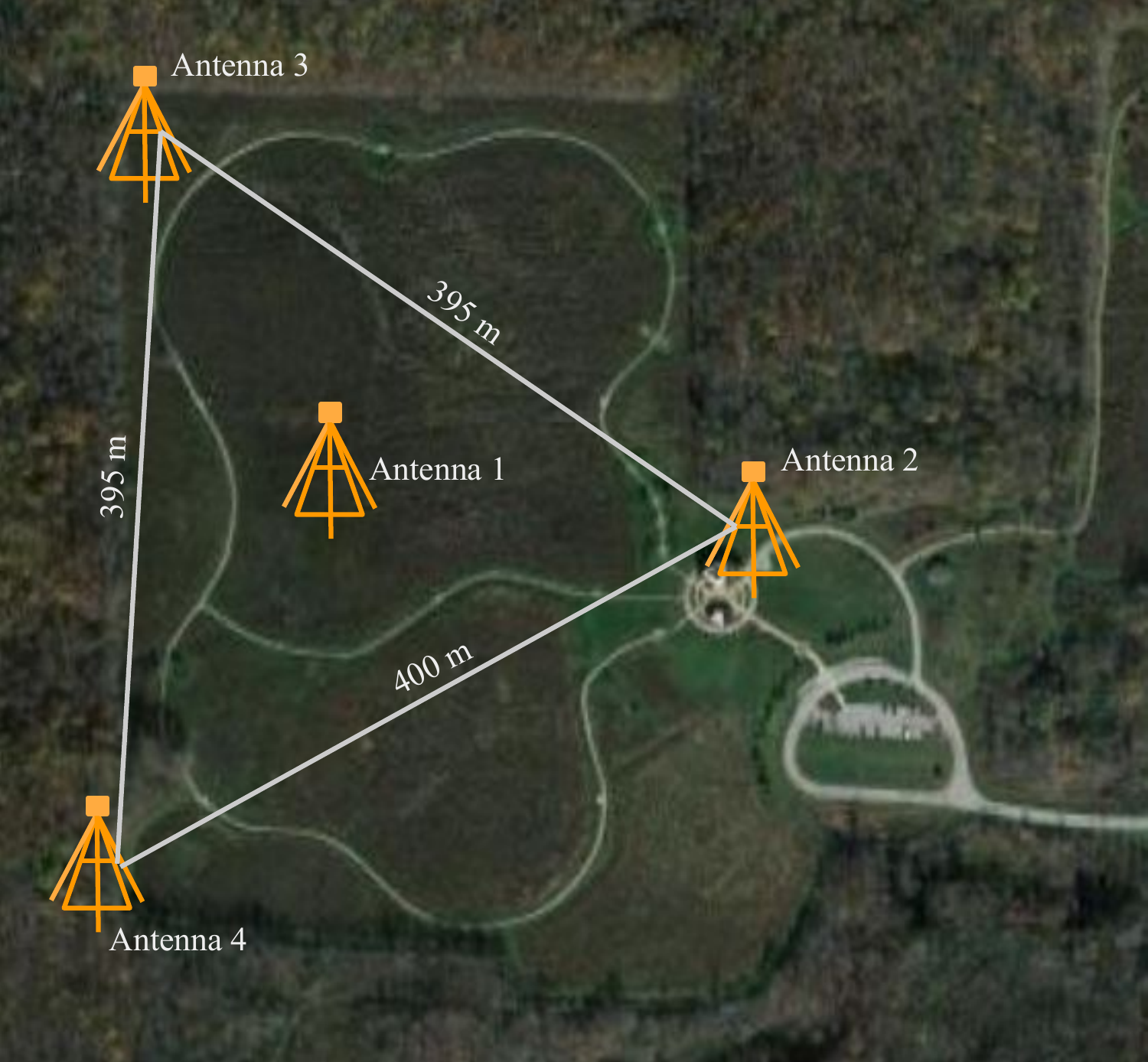}
       \caption{\label{fig:DLITE_diagram_maps} Aerial site map of Observatory Park with locations of each DLITE antenna indicated by orange diagrammatic antennas. DLITE arrays are composed of four dipole antennas. Due to the small number of elements in the array, the optimal configuration for the DLITE array is an equilateral triangle with one antenna at each point and a fourth in the middle. This configuration maximizes the number of unique baselines for a four-element array which increases imaging and source tracking capabilities. In order to resolve the brightest radio objects in the sky, the three corner antennas should be separated by at least 250 meters, with the optimal separation being between 350 and 500 meters. The separations between the three outer antennas of DLITE Ohio are highlighted, with all separations being around 400 meters based on Google Maps measurements \citep{DLITE}. Antenna 2 (41°35'07.2"N 81°04'49.4"W) is the closest to the Observatory Park Science Center and parking lot, Antenna 3 (41°35'14.0"N 81°05'03.9"W) is located in the far North Western corner of the park, Antenna 4 (41°35'01.3"N 81°05'04.9"W) is located in the far South Western corner of the park, and Antenna 1 (41°35'08.0"N 81°04'59.4"W) is centrally located in the park. Each antenna location has a unique radio frequency interference (RFI) environment, with Antenna 2 seeing on average more RFI than the other three antennas, likely due to a transformer box close by. Observatory Park is well shielded on all sides from RFI due to the mature forest surrounding the telescope site \citep{trees}.  }

  \end{figure}

Optimally, the antennas in a DLITE array are arranged in an equilateral triangle: three antennas at the vertices separated by between 200 and 500 meters with a fourth antenna in the center of the pattern. This configuration maximizes the number of unique baselines for a four-element array, and increases imaging and source tracking capabilities. \citep{DLITE} As Observatory Park is relatively flat and has a large, uninterrupted area of grassland, DLITE Ohio is very close to the optimal equilateral triangle setup for a DLITE array as seen in Figure \ref{fig:DLITE_diagram_maps}, with each of the corner antennas separated by between 390 and 410 meters. Baselines in this range are long enough to resolve the A-Team sources, but short enough to still have them appear point-like, thus enabling observations of the scintillation of the ionosphere. The separations between the three outer antennas at DLITE Ohio are roughly 400 meters, which effectively suppresses extended galactic emission. The effective resolution on the sky for a baseline $B$ scales as $= B^{-1}$. The minimum separation of the A-Team sources is 20$\degree$, thus for one baseline of 400 meters and frequency bandwidth of 8.33\,MHz, our resolution of 5$\degree$ is sufficient to resolve the sources \citep{DLITE}.

\subsection{DLITE System Design Description}

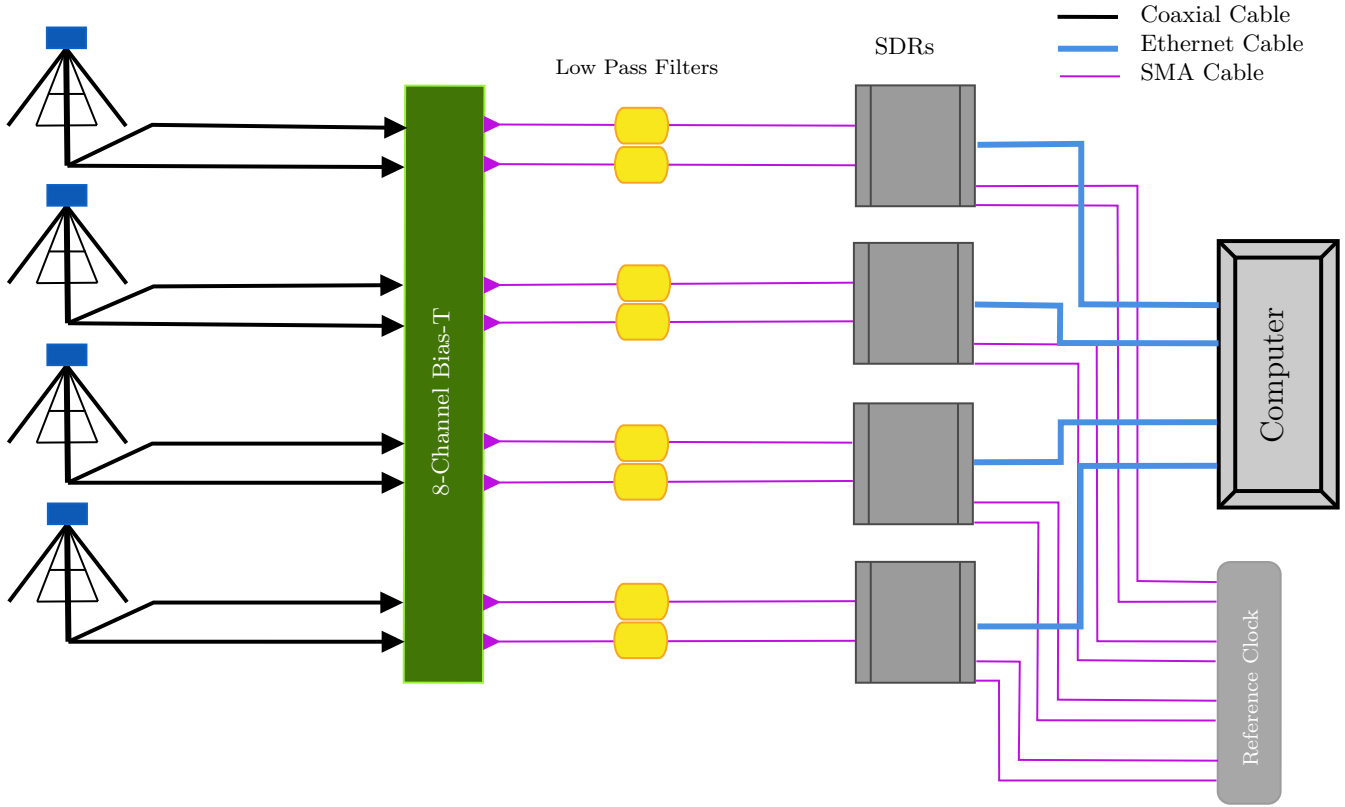
\begin{figure*}[t!]
    \centering
    \input{DLITE_system_diagram}
    \caption{System diagram for DLITE Ohio, recreated from Figure 1 in \cite{DLITE}. Specific information about individual components can be found in Table \ref{tab:parts}.}
    \label{fig:dlite_matic}
\end{figure*}

The DLITE system implemented at Observatory Park closely follows the recommendations presented in \cite{DLITE}. Figure \ref{fig:dlite_matic}, modeled after a diagram from the original paper, details the data flow through the system's hardware. The DLITE digital backend is assembled within a Pelican case that contains a 14U rack, which also houses a DC power supply.

The antennas of DLITE Ohio are connected to the backend via short lengths of LMR-240 within the antenna masts and lengths of LMR-400 coaxial cable comprising the majority of the baseline \citep{LWA_memo}. The coaxial cable is then connected to an 8-channel bias tee module custom-made by the Naval Research Laboratory for the DLITE system \citep{2012PASP..124.1090H}, although off-the-shelf bias tee units will also suffice. The signal is then routed through low-pass filters with a cutoff at 50\,MHz and into off-the-shelf software-defined radios (SDRs). These SDRs are both connected to a 10\,MHz reference clock and the backend processing computer. For a complete list, including part numbers and prices, of materials used to commission DLITE Ohio, see Table \ref{tab:parts} in Appendix C.

The DLITE correlator is written with GNU Radio to enable communication with the SDRs, with all processing blocks needed for basic cross-correlating natively available within the environment. Details about the correlator and a GNURadio flowchart can be found in \cite{DLITE}.

\section{Commissioning}

\label{sec: com}

The commissioning for the DLITE array at Observatory Park was student-led and completed over the six months leading to the 2024 North American total solar eclipse. Marking the antennas locations and radio frequency interference (RFI) environment tests were completed in September 2023, and the telescope was online by March 23, 2024. We are working toward DLITE Ohio becoming a long-term fixture at Observatory Park, with the goal of being able to continue to probe the ionosphere and observe the Sun for many years.

Observatory Park is located $\sim$50 kilometers from Cleveland, Ohio, in the rural farming town of Montville. The site is relatively well isolated from RFI, allowing for low-frequency observations. This is largely due to being away from any population center and surrounded by a mature hardwood forest \citep{trees}. There is persistent, albeit weak, RFI close to the observatory's multi-purpose building, likely due to the nearby electrical transformer box. Additionally, there is a strong and persistent RFI of an unknown source in the Y (East-West) polarization of all four antennas. 

The commissioning of DLITE Ohio was performed entirely by students, and comprised of the following steps:

\begin{enumerate}
    \item \textbf{RFI measurements}: Undergraduate students used a handheld spectrum analyzer to measure the radio frequency environment for each antenna location. We found minimal interference in the 35--45\,MHz frequency band from significant noise sources such as powerlines. See Appendix B for an in-depth discussion of these site survey measurements and further noise environment characterization completed after the installation of the full array.

    \item \textbf{Antenna installation}: DLITE arrays employ antennas originally developed for the LWA. These antennas are composed of four aluminum vanes connected via fiberglass support structures and a head unit used to house the front-end electronics (FEEs) that sit on an aluminum anchor mast which is manually driven into the ground (see Fig.~\ref{fig:DLITE_pic}). Students assembled these antennas, secured them on the installed anchor masts, and aligned the antennas vanes with the North-South and East-West polarization \citep{LWA_memo}.

\begin{figure}[h!]
    \centering
         \includegraphics[width =0.9 \linewidth]{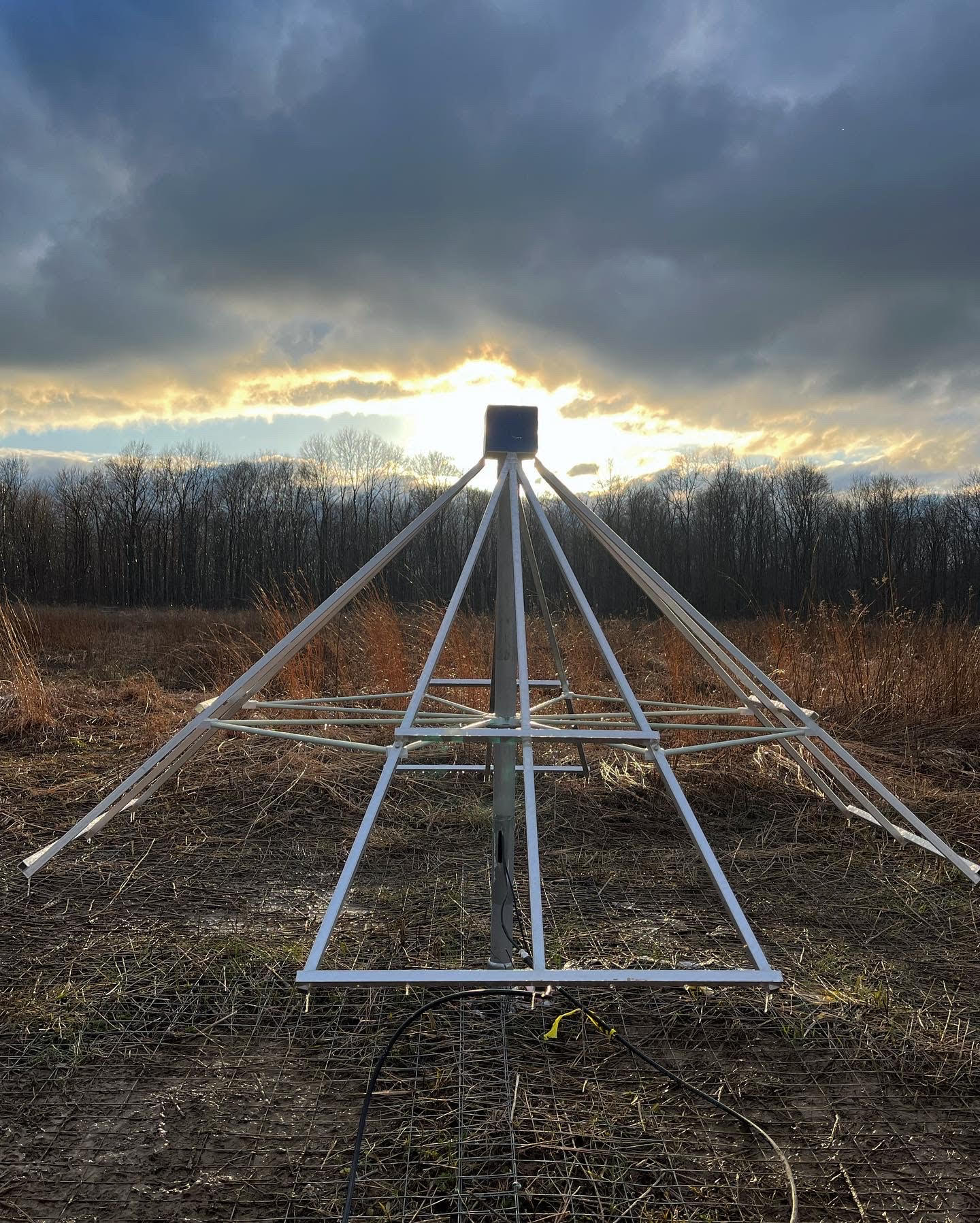}
        \caption{Image of the central DLITE antenna at Observatory Park, Ohio.}
        \label{fig:DLITE_pic}
       
  \end{figure}
    
    \item \textbf{Backend installation and single baseline first light}: The backend electronics for DLITE are mostly comprised of off-the-shelf parts with a detailed bill of materials given as Table \ref{tab:parts} in Appendix C. These backend hardware components, when coupled with software developed for the DLITE project in GNURadio, allow for swift commissioning with little additional user input. The backend for DLITE Ohio was installed on March 11, 2024 and brought online, enabling remote access. The first light for the array was achieved by laying coaxial cable along our shortest baseline, allowing for system tests to be run and the RFI environment to be further characterized (see Appendix B). Additionally, multiple SRBs were observed using this single baseline. 
    
    \item \textbf{Cable laying and full-array first light}: DLITE currently employs two types of coaxial cable: LMR-400 for the long baseline runs and LMR-240 to run inside the antenna masts and connect to the FEEs. Over 2.5\,km of LMR-400, a low-loss, dual-shielded telecommunications cable, was used for this project. Students laid out cables to our three longer baselines, learned to solder on cable end connectors, and searched for and fixed problems within the system. The first light for the full array was achieved on March 23, 2024. This allowed for more than two weeks of software troubleshooting and data-taking leading up to the eclipse. 
\end{enumerate}

The commissioning for the eclipse, not including trenching for permanent coaxial cable installment, was completed over the course of six on-site work days at Observatory Park over six months. Additionally, a prime focus of this work was student skill development, with most students coming into the project with little background in radio instrumentation. Therefore, assuming procurement of all needed hardware and not including trenching for permanent installation, this project has shown that the DLITE system can be deployed and fully operational in the span of a week with very little background experience required.

\section{Observations during the April 2024 total solar eclipse}

\label{sec: obs}

A total solar eclipse was observed by DLITE Ohio on April 8, 2024, as part of a multi-day campaign to both observe the solar corona and probe the ionospheric response to the eclipse. The optical eclipse began at 18:00 UTC and lasted until 20:29 UTC with totality lasting for around 3 and a half minutes from 19:14 UTC and 19:18 UTC.\footnote{\url{https://eclipse-explorer.smce.nasa.gov/}}

\subsection{Solar activity during the time of the Eclipse}

\begin{figure}[h!]
    \centering
         \includegraphics[width = 1.0\linewidth]{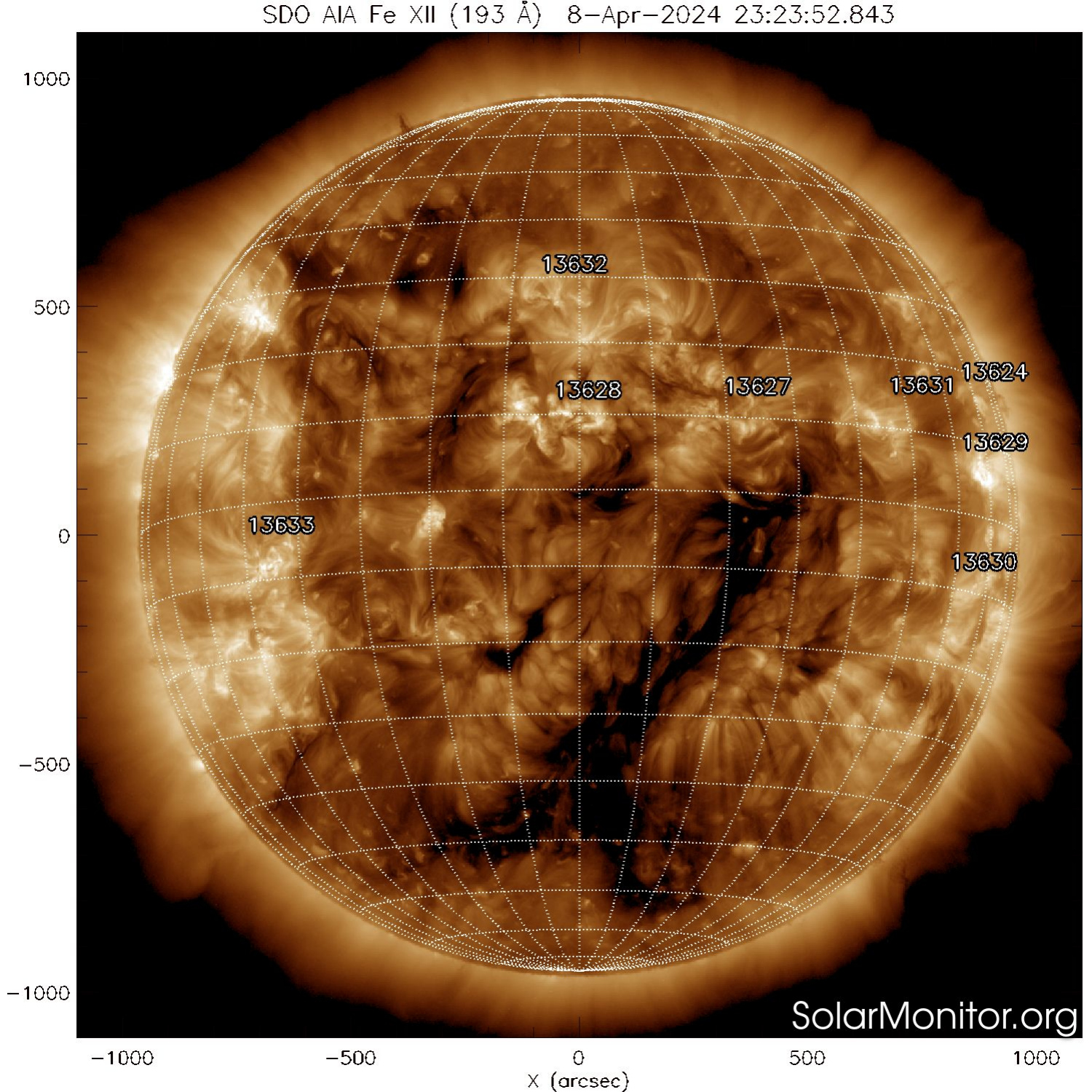}
       \caption{\label{fig:SDO} SDO/AIA image at 193\,$\mathrm{\AA}$ from the day of the eclipse. NOAA active regions are highlighted. Credit: SDO/AIA \cite{SDO, AIA}.}
       
  \end{figure}

\begin{figure*}[p!]
    \centering
         \includegraphics[width = 1.0\linewidth]{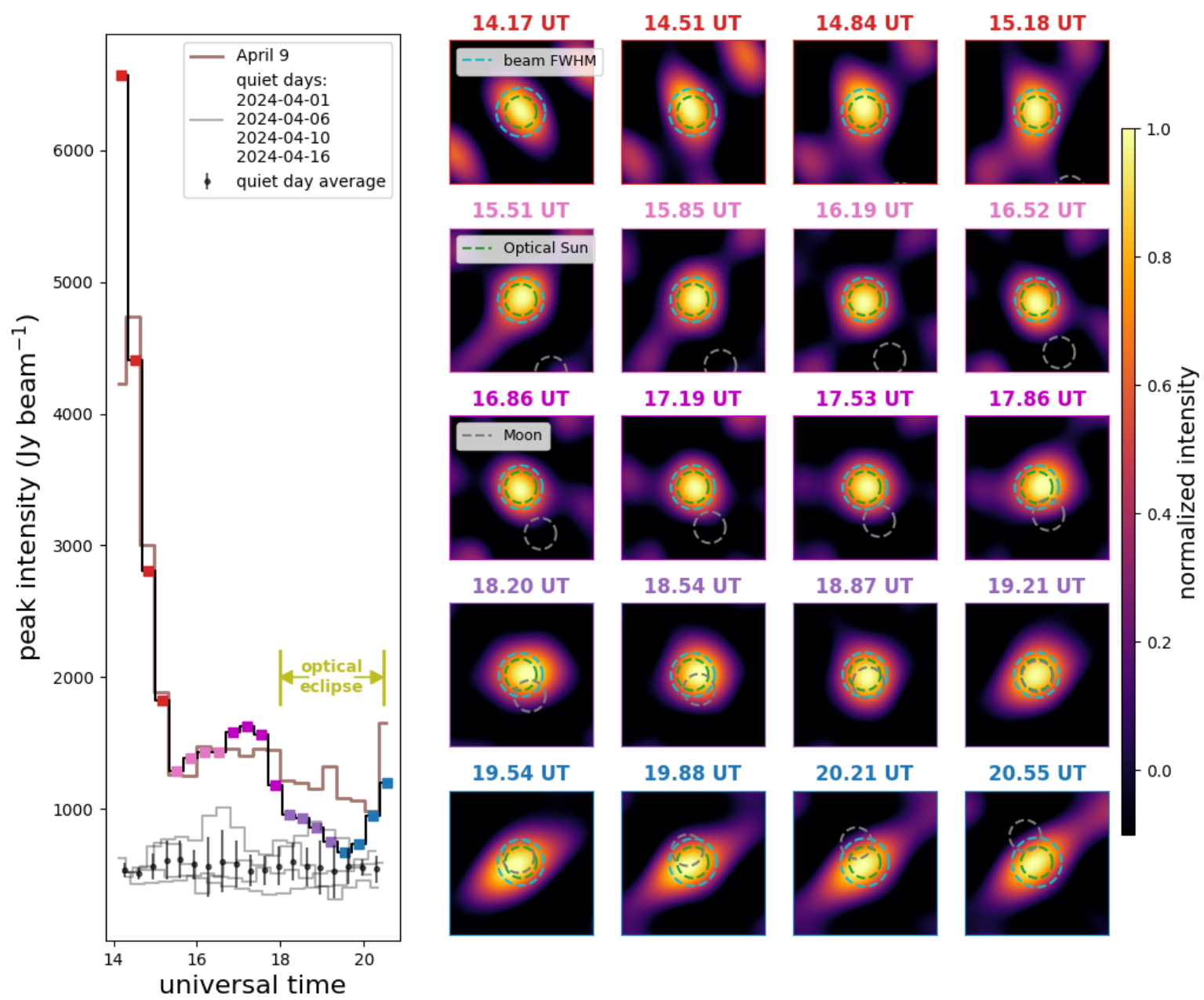}
       \caption{\label{fig:images} Visibilities and peak intensities for DLITE Ohio observations on the day of the total solar eclipse given every 20 minutes. The location and size of the optical Sun are indicated with a green dashed line ($\sim$0.5$\degree$), which is smaller than DLITE's beam at FWHM on the sky ($\sim$0.6$\degree$), indicated in light blue. The movement of the Moon is indicated with a grey dashed circle, with the peak of the optical eclipse seen in the panel labeled 19.21 UT. The peak intensity for each postage stamp image is shown in the subplot to the left, with the time-step markers color-coded in the same manner as the time stamps on the right, with a notable dip occurring around the time of the optical ellipse (indicated by the green bars). We note that the features around the edges of the images are sidelobes, features caused by short time-scale fluctuations in the solar intensity (see Appendix A), from which the solar radius is sufficiently isolated. The subplot on the left also includes peak intensities from observations on the day after the eclipse, April 9, 2024, indicated by the brown line, along with the average peak intensities of several quiet days leading up to and after the eclipse. During solar maximum, the intensity of the sun is highly variable, but the behavior on the day of the eclipse and the day after closely align at these frequencies. The only significant deviation between the two days is the decrease in intensity corresponding with the time of the optical eclipse on April 8, which is not present in the data from April 9, 2024. We note here that there is evidence of a strong southward gradient in the total electron count (TEC) of the ionosphere (see section \ref{techs}) on both the 8th and 9th during the times that Cas and Cyg A are highest in the sky between 14:00 - 15:00 UTC, which correspond to the times of the large peaks in the average intensity seen on April 8th and 9th. Strong gradients in TEC can cause confusion with other sources that are better mitigated as the A-Team sources set, which results in more accurate isolation of solar radiation. Using quite days before and after the eclipse, we can more accurately quantify and isolate our instrument’s response to solar radiation. The dip seen with DLITE-Ohio that corresponds with the time of the optical eclipse on April 8th, 2024 is statically significant when compared to average deviations in solar emission on quiet days at these frequencies. We also note that after the end of the optical eclipse, the Sun dips below 30$\degree$ elevation. Both the sensitivity of the LWA-style antennas used for DLITE and the uncertainties of the antenna gain corrections become significantly less reliable for sources below 30$\degree$ elevation, therefore further data points are not included (sections 2.3 and 3.0 in \cite{Dowell_2017} for further discussion on source elevation dependence).}
       
  \end{figure*}

There were several active regions on the Sun the day of the eclipse (NOAA 13628, 13629, 13630, 13631, 13632, 13633, 13624, 13627), although there were no significant flares, as defined by NOAA in terms of peak emission in the 0.1 – 0.8\,nm spectral band (soft x-rays) of greater than $10^{-8}\,\mathrm{\frac{W}{m^2}}$. We do note that in our observations (see Figure \ref{fig:images} in the next section) there is an uncharacteristic spike in low-frequency radio solar emission before the eclipse, which may be evidence of a minor solar flare due to the relatively low flux. There were also several coronal holes (CH1, CH2, CH3, CH4, CH5) seen using CHIMERA \citep{Garton_2018}.  Additionally, a significant solar prominence was visible using the naked eye during totality as seen from Observatory Park.  A Solar Dynamics Observatory/Atmospheric Imaging Assembly (SDO/AIA\textit{}) \citep{SDO, AIA} image at 193\,$\mathrm{\AA}$ taken at 23:23 UTC on the day of the eclipse is presented in Figure \ref{fig:SDO}, with both the regions of activity and the solar prominence visible. These features may affect the peak intensity of the solar corona and the overall brightness temperature as observed by DLITE Ohio as we are unable to resolve these features independently. 

\subsection{DLITE Techniques for the Eclipse}  
\label{techs}

For solar observations during the eclipse, standard synthesis imaging techniques were employed to combine data from all antennas in the array. During the eclipse, there were significant noise sources in all four of our Y (East-West) polarization baselines. Thus, these images are composed of data from all four X (North-South) polarization baselines, or six baselines in total, which allows for semi-resolved imaging of the Sun with $0.6\,\degree$ resolution. As DLITE requires some system specific approaches to synthesis imagining, our process is described in more detail in Appendix A. 

Additionally, the Sun is far from the brightest object in the sky at these frequencies. Radio emissions from the bright A-Team sources must be removed from our images to ensure they are not contributing to our measurements of solar intensity. A peeling technique, which is discussed in Appendix A, is used to fully remove both emissions from Cassiopeia A (our ionospheric target during the eclipse) and ionospheric scintillation toward Cas A caused by structures in the ionosphere comparable to the Fresnel scale ($\sim$2\,km at 35\,MHz). Although these features must be removed to better image the Sun, they will be presented in an upcoming paper as observing bright radio sources through the path of totality allows for investigation into the ionospheric response to the eclipse.

 Figure \ref{fig:images} shows the calibrated and peeled observations of DLITE Ohio leading up to, during, and after the peak of the eclipse (see Appendix A for details). The radio Sun is centered in each subplot of the grid to the right of the plot, with the disk of the Sun isolated from sidelobe artifacts visible around the borders of the subplots. The footprint of the optical Sun is highlighted by the green dashed circle, with the blue dashed circle indicating the footprint of the FWHM of our beam. While the optical Sun is smaller than our beam at FWHM, the radio Sun is observed to extend further. The Moon is indicated by a grey dashed circle, first emerging at the bottom right corner of the subplot at 15.18 UT. Its position was calculated using the Jet Propulsion Laboratories Horizons System \footnote{\url{https://ssd.jpl.nasa.gov/horizons/}}. The second major subplot within this figure details the average peak intensity of the Sun over the course of our observation. As noted previously, there appears to be a spike in the intensity at 14.17 UT, potentially from a minor solar flare, which dims until stabilizing at around 15.51 UT, over two hours before the start of the eclipse as seen in Montville, Ohio.

 As the optical eclipse began in Montville at 18:00 UT, we observed the onset of a corresponding decrease in radio emission as seen with DLITE Ohio, illustrated in the left-hand subplot in Figure \ref{fig:images}. The intensity of the Sun continued to decrease as totality approached; however, the lowest decrease was not reached until $\sim$15 minutes after the end of optical totality. Radio emission is not uniform in the solar corona \citep{LoFar}; thus, during the eclipse, there was likely a significant region of emission in the upper left corner of the middle corona which was covered following optical totality.

Although solar intensity during solar maximum is highly variable, the peak intensities recorded from the day after the eclipse (April 9, 2024) closely correspond to our data on the day of the eclipse (see Figure \ref{fig:images}). The only significant deviation between these two days of observations occurs during the time of the optical eclipse, where there is a notable decrease in solar intensity on the day of the eclipse when compared to the same time range the day after the eclipse. Therefore, the peak solar intensities on April 9, 2024 are used as a proxy for the average expected solar emission during the time the eclipse occurred on April 8, 2024. Although a full analysis of the ionospheric conditions leading up to and during the eclipse is outside the scope of this paper, we note here that there is evidence of a strong southward gradient in the total electron count (TEC) of the ionosphere from MIT’s Madrigal database \footnote{\url{ http://cedar.openmadrigal.org/}} on both the 8th and 9th during the times that Cas and Cyg A are highest in the sky. Strong gradients in TEC can be associated with small-scale irregularities which increase scintillation in the direction of A-TEAM sources \citep{2023SpWea..2103442H}. These irregularities can cause confusion with other sources such as the Sun, rendering our peeling techniques used to isolate solar emission less effective as the model visibilities generated during peeling do not accurately account for short timescale fluctuations of A-Team sidelobe emission (see Appendix A). These gradients are the likely cause of the large peaks between 14:00 - 15:00 UTC on April 8th and 9th. However, as the A-TEAM sources set, the confusion is better mitigated by our peeling techniques. This combined with the TEC gradients stabilizing resulted in more accurate isolation of solar radiation in the hours leading up to and during the eclipse. There were several quiet days, both in terms of solar and ionospheric activity, within the two weeks before and after the eclipse. Using these days, we can more accurately quantify and isolate our instrument’s response to solar radiation. During these quiet days, the average deviation from the mean solar intensity is around 140 Jy. Therefore, the dip seen with DLITE-Ohio that corresponds with the time of the optical eclipse on April 8th, 2024 is statically significant when compared to average deviations in solar emission on quiet days at these frequencies. We also note that the day before the eclipse, April 7th, is not included in this analysis, as it was a very active day both in terms of solar and ionospheric activity.

The solar intensity at 42 MHz during the peak of the optical eclipse at 19.21 on April 8 drops to $\sim$751\,Jy $\rm beam^{-1}$, while the peak solar intensity at 19.21 UT on April 9 is $\sim$1324\,Jy $\rm beam^{-1}$. This corresponds to a 43$\%$ decrease in the low-frequency radio emission centered at 42\,MHz as observed from the path of totality. This decrease would be expected if low-frequency solar radiation originates from the middle corona at an average of $\sim$1.5 optical solar radii, assuming solar emission in these frequencies is on average circular and consistent across the observed solar disk. This finding corroborates models based on plasma emission as the main mechanism for low-frequency radio emission in the middle corona \citep{West_2023, Gary2005}. Although previous observations of the quiet Sun with LOFAR show that solar emission is on average neither circular nor consistent at these frequencies, with the horizontal (E-W) width being significantly larger than the vertical (N-S) width \citep{LoFar, Zhang_2022}, the projected beam size of the DLITE system is too large to resolve these features reliably.

Additionally, these observations allow us to measure the brightness temperature of the Sun directly via the use of the Rayleigh-Jeans Law to convert our measured intensities to brightness temperatures. Leading up to the eclipse, the measured peak solar intensity averaged around 1440\,Jy $\rm beam^{-1}$, corresponding to a brightness temperature of around $3.09 \times 10^{5}\,\mathrm{K}$. This measurement agrees well with those for quiet regions of the Sun between 29 and 50\,MHz \citep{1999spro.proc...11L, 1987SoPh..111..419W}. Although these measurements with DLITE Ohio were taken going into solar maximum, it was on a relatively quiet day (see Section \ref{sec: obs}) and based upon the peak intensity of our beam on the Sun versus resolved quiet or active regions as was done in these previous works.

\section{DLITE TV: Outreach During the Eclipse}
\label{sec: outreach}

\begin{figure}[h!]
    \centering
         \includegraphics[width = 1.0\linewidth]{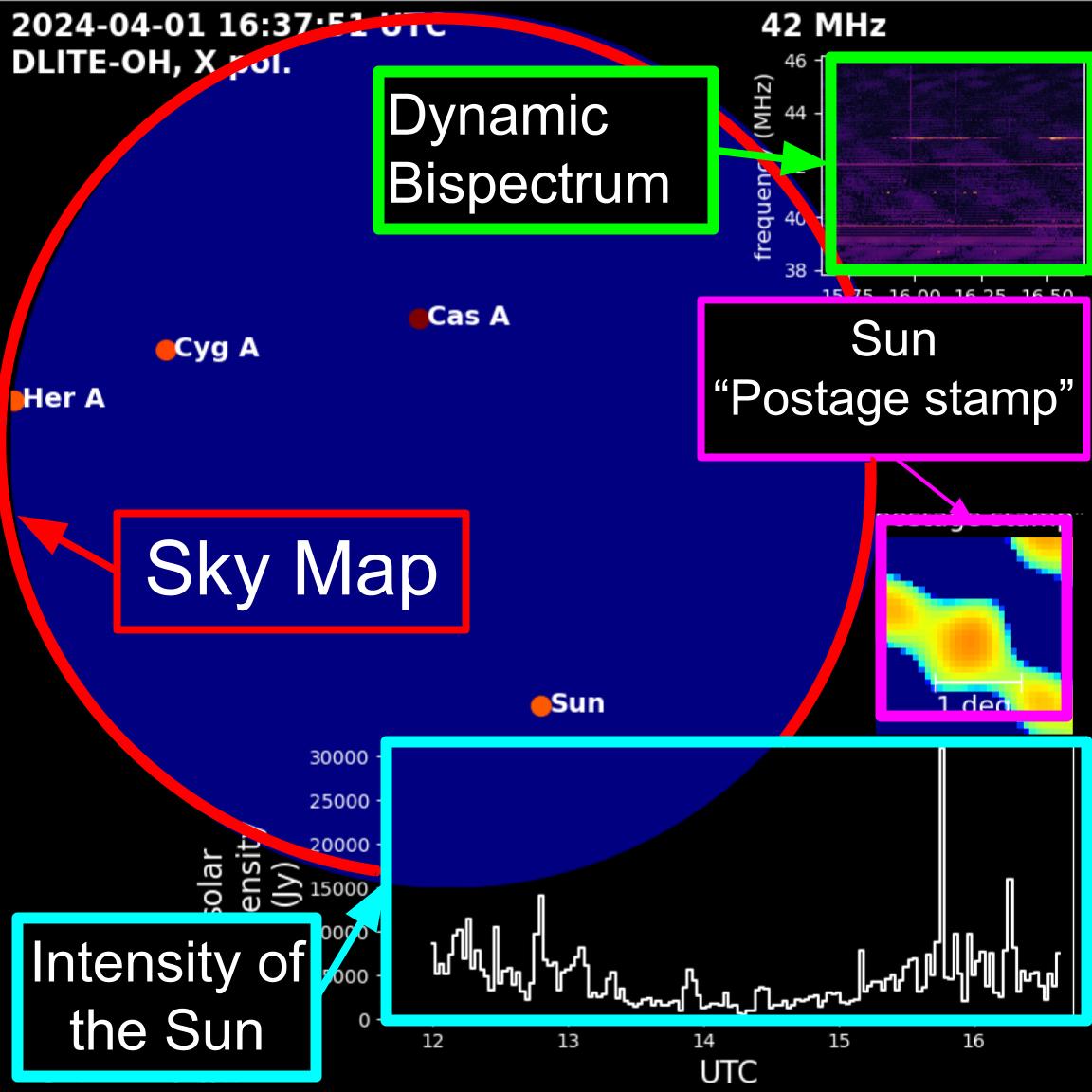}
       \caption{\label{fig:DLITE_TV} Labeled example of our live stream data viewer. Visitors to our website during the eclipse saw a close to real-time data stream of the eclipse in radio emission taken with DLITE Ohio, including a live sky map showing the location and brightness of the Sun and the A-Team sources, a ``postage stamp'' image of the Sun, a plot of the intensity of the Sun vs time, and a dynamic bispectrum. This plot was updated every $\sim$4 minutes during the eclipse. Along with descriptions of each of these subplots, the website also hosted explanations of the DLITE project and our science goals.}
       
  \end{figure}

This project also included the development of a close-to-real-time data live stream inspired by LWA-TV \footnote{\url{https://leo.phys.unm.edu/~lwa/lwatv.html}} and a project website to both host the live stream and provide details about our science goals.\footnote{\url{https://olivia-r-young.github.io/}} Figure \ref{fig:DLITE_TV} is a snapshot of the live data stream. 

During the eclipse, there were over 900 views to our data live stream from 30 states and 11 counties, allowing many people outside Observatory Park to experience the eclipse in radio emission from our telescope. These website analytics were provided through the open source service hosted on GitHub, GoatCounter \citep{goatcounter}.

\subsection{Open source data processing scripts}

An eventual goal of the DLITE project is for many DLITE arrays to be in operation around the world, built by universities, laboratories, and individuals. The student-led construction and commissioning of DLITE Ohio proves that the DLITE system can be deployed quickly without the need for an extensive background in radio instrumentation.

To further increase the accessibility to the project for students, an open-source data processing Google Colab notebook was developed, allowing for not only ease of physical construction but also streamlining the path toward using the array to study the ionosphere and the Sun. This processing notebook walks through the generation of visibility plots for one baseline, source flagging, and the creation of a video of the visibilities for the full observation, essentially allowing users to recreate the sky maps featured in our DLITE TV live stream. 

This notebook can be found at \url{https://zenodo.org/doi/10.5281/zenodo.13286741}.

\section{Future Outlook}

 In this paper we described the student-led commissioning of a DLITE radio telescope at Observatory Park in Montville, Ohio, to observe the solar corona during the 2024 total solar eclipse. We have detailed the first low-frequency radio observations taken in the path of a total solar eclipse during a solar maximum and found the radius of emission at 42\,MHz to agree with expectations based on plasma emission in the middle corona \citep{West_2023, Gary2005}. We have also shown the efficacy of using DLITE arrays to observe rare and localized events, showcasing the flexible deployability of this telescope design and its potential as both a scientific instrument and a teaching tool.

We have shown that DLITE Ohio gives insight into both astrophysical and ionospheric phenomena and can be a powerful tool for teaching students. However, true insight into astrophysical phenomena such as coronal emission or insight into the large-scale structures of the ionosphere at these frequencies requires many stations with large separations as has been exhibited by the success of arrays such as the LWA and LOFAR. During the eclipse, two additional DLITE stations were taking data from different locations along the path of the eclipse, although neither were in totality. A fully operational array in Pomonkey, Maryland, experienced an 85.7$\%$ optical eclipse, and a partially operational array in Socorro, New Mexico, a 75$\%$ optical eclipse \footnote{\url{https://eclipse-explorer.smce.nasa.gov/}}. The results of combining these data will contribute to a larger collaboration paper on solar activity and coronal observations using data from across many low-frequency radio telescopes that observed the eclipse. DLITE Ohio will still be the only low-frequency radio interferometer to have observed from the path of totality, making its contribution unique. 

Additionally, data taken by DLITE Ohio during the eclipse will contribute to a larger DLITE collaboration paper on the ionospheric response to the eclipse. Since DLITE was specifically designed to use the bright A-TEAM sources as probes of the ionosphere, using all three DLITE stations to probe different lines of sight through the path of the eclipse will provide a new perspective on how the ionosphere responds to an eclipse in different geographical regions. 


\section{Acknowledgments}

\emph{Author Contributions.} O.R.Y., T.E.D., and J.F.H. conceived, directed, and executed the project, performed field and equipment work, and developed and edited the manuscript. L.P.D. performed field and equipment work, software development, and edited the manuscript. M.T.M. and S.V.S.F. conceived the project and performed initial feasibility analyses.
C.M., E.B., C.J., J.L., G.M., A.P., J.P., P.R., E.S., B.B., A.B.H., A.Z.H., P.L., W.L., M.M., and N.H. conducted field work, testing, and deployment of equipment at Observatory Park. C.M., in an especially dedicated fashion, contributed on-site work during every site visit to build the telescope, conducted testing, made deployment recommendations, and garnered support and excitement for the project within the local community. M.Z. extensively edited the manuscript and analysis.

We wish to acknowledge our funding and support sources for this work which include the American Astronomical Society's Jay M. Pasachoff Solar Eclipse Mini-Grants Program, Hillsdale Collage Division of Natural Sciences, Hillsdale College Dept. of Physics, Hillsdale College LAUREATES funding, and the Naval Research Laboratory. Contributions ranged from funds for buying cables to the donation of our backend electronics.

We thank Gregory B. Taylor and Jayce Dowell for hosting T.E.D., O.Y, and Hillsdale students at the University of New Mexico on training trips to the Long-Wavelength Array Swarm mini-station at the North Arm of the Very Large Array. We thank Hillsdale College for continued support of the on-campus Low-Frequency All-Sky Monitor V radio telescope, with which students were also trained. We are grateful to Douglas Hamilton for the donation of a handheld spectrum analyzer to the Hillsdale College Dept. of Physics. We thank Hillsdale College students Evan Anthopoulos, Riley Hamilton, Nathan Sibert, and Liam Swick for their off-site preparatory work on DLITE equipment. We also thank Burns Industries for rapid delivery of junction boxes in preparation for the eclipse.

We are especially grateful to Observatory Park and the Geauga Parks District for hosting our DLITE array, as well as providing advancement funds for the purchase of our coaxial cable. We also thank Petru Fodor at Cleveland State University for the idea of collaborating with Observatory Park as a DLITE site. We are additionally grateful to Martin Joyce for his time and his lending of equipment for the site construction.

O.Y. is supported by the National Science Foundation Graduate Research Fellowship under Grant No. DGE-2139292. Additionally, she would like to thank the National Radio Astronomy Observatory for her current position as a visiting graduate student.

T.E.D.'s preparatory summer and sabbatical work was supported by an NSF Astronomy and Astrophysics grant (AAG) award number 2009468 and, along with the Student Teams of Astrophysics ResearcherS (STARS) program, was supported by the NANOGrav NSF Physics Frontiers Center under award number 2020265. M.T.L. also graciously acknowledges support from these two awards. L.D. is supported by NSF
award PHY-191259.

Basic research at the Naval Research Laboratory (NRL) is supported by 6.1 base funding.

\emph{Dedication.} This paper is dedicated to the memory of co-author Emerson Sigtryggsson, and to her family and friends. Emerson was a Hillsdale College physics major who lost a brief but valiant fight against cancer. Her enthusiasm as a LAUREATES summer student on this project shined, and we would not have completed DLITE-OH without her perseverance and hard work.

\bibliography{sample631}{}
\bibliographystyle{aasjournal}

\appendix

\section{Image Processing for the 2024 Total Solar Eclipse with the DLITE System}
To constrain the size of the solar emission region at 42 MHz, we developed algorithms to synthesize the entire DLITE bandwidth among all six baselines to generate broadband images of the Sun.  These are a combination of standard synthesis imaging methods described by, e.g., \citet{thompson91} and a DLITE-specific ionospheric analysis algorithm detailed by \citet{DLITE}.  Because the Sun is typically significantly fainter than some A-Team sources such as Cyg A and Cas A, which are both around 22,000 Jy at 38 MHz, and the entire sky is visible to the DLITE antennas, it was necessary to implement a scheme similar to peeling \citep[see, e.g.,][]{intema09} to mitigate sidelobes from those bright sources within any image centered on the Sun.  Within peeling, iterations of self-calibration and the CLEAN algorithm are used to solve for instrumental and direction-dependent ionospheric phase errors toward a bright source within the field of view.  Model visibilities are then generated from the CLEAN components, which are corrupted with the determined phase errors so that the source can be subtracted from the original observed visibilities to mitigate its impact on imaging the rest of the field of view.

With only four antennas and six baselines, self-calibration would be poorly constrained in the DLITE case.  However, as shown by \citet{DLITE}, it is possible to iteratively solve for the ionospheric position errors and intensity variations of all visible bright sources, the former of which have been used to study medium scale ($\sim$50-200 km) ionospheric disturbances \citep{helmboldt22}.  The position shifts can be recast as phase gradients over the array, which in turn can be converted to differential phases between the antennas of each baseline.  Ionosphere-corrupted model visibilities of the bright sources are then constructed using these differential phases and the determined intensity variations.  In practice, a (complex) linear combination of these model visibilities (plus a constant) is fit to and subtracted from the visibilities of each baseline to ensure that a reasonably constrained combined model is used.

Following this peeling step, a time period when the Sun was visible to the array is selected, and the peeled visibilities are fringe-stopped toward the Sun, that is, the data for each baseline are corrected for the geometric and instrumental delay for the known position of the sun \cite[see][for a description of how the instrumental delays are determined]{DLITE}.  To ensure any remaining gain errors are accounted for, the visibilities for each baseline are divided by the mean over all times and frequencies.  While it is more typical to use self-calibration to do this for what should be antenna-based and not baseline-based errors, as noted above, the small number of antennas precludes this.  The risk of this baseline-based approach is that it can potentially force the data to conform to point source-like visibilities.  However, this is mitigated by using the average over a relatively large range of times ($\sim$1 hour or more) and all frequencies.  We also note that the Sun will only be marginally resolved by DLITE, which also mitigates this risk.

\begin{figure}[h!]
    \centering
         \includegraphics[width = 1.0\linewidth]{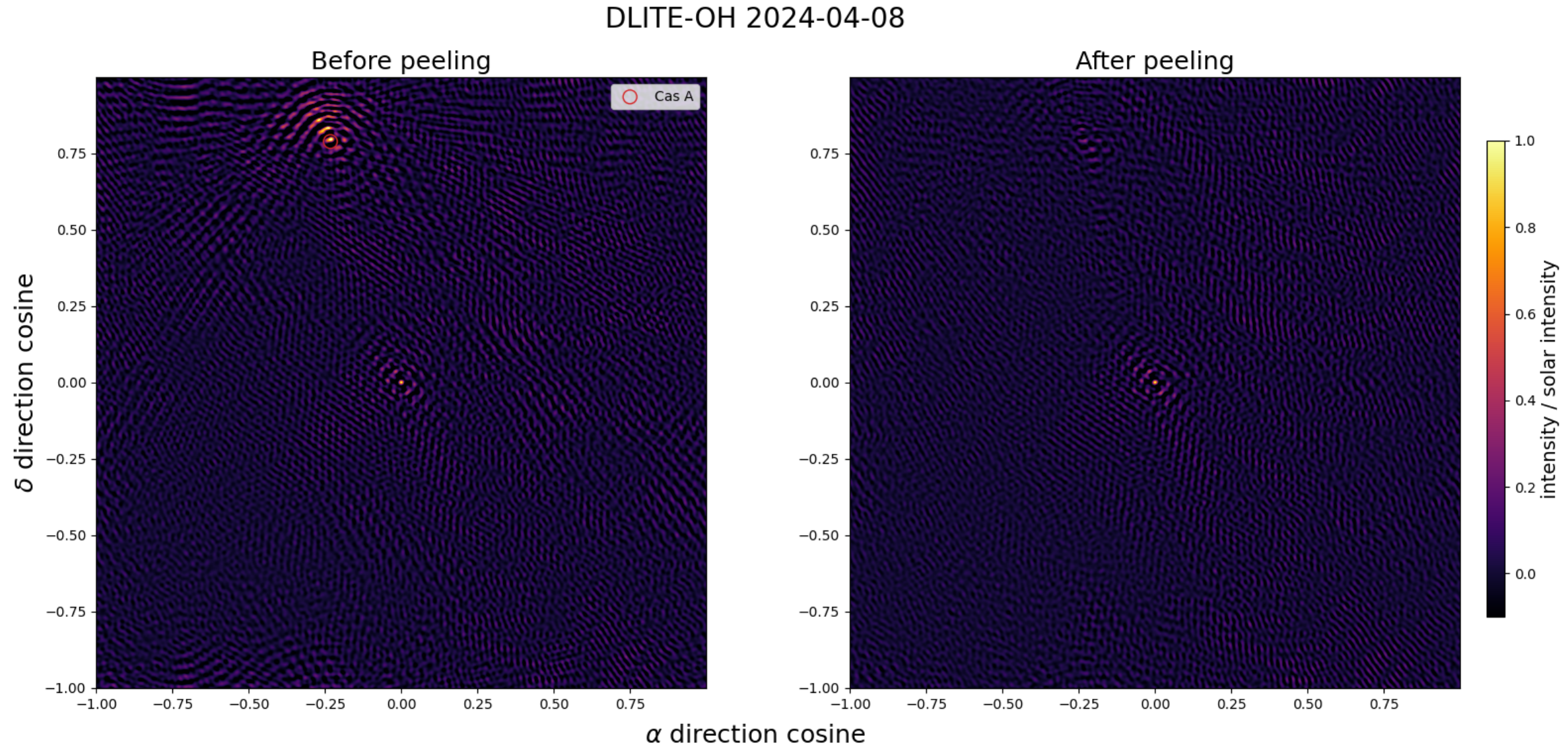}
       \caption{\label{fig:peel_1} For all times when the Sun was above $30^\circ$ elevation ($\sim$7.5 hours of data), images centered on the Sun before (left panel) and after (right panel) the peeling process described in Appendix A was applied. The location of Cas A is indicated in the left panel by a red circle, and the Sun is the single bright point source centered in both plots.}
       
  \end{figure}

The peeled, fringe-stopped, and gain-corrected visibilities are then gridded and summed within the $u,v$-plane, in our case using the NumPy\footnote{\url{https://numpy.org}} function {\tt histogram2d}.  This is the equivalent of natural weighting.  With so few baselines, a more complicated weighting scheme is not advisable.  The gridded visibilities are then Fourier transformed (with the NumPy {\tt fft.fft2} function) to produce an image.  Note, this imaging method assumes the antennas are coplanar, i.e., that they are all at the same altitude.  While precise antenna altitudes are not known, they certainly differ by less than a wavelength ($\sim$7\, m).  Fig.\ \ref{fig:peel_1} shows an image from the date of the eclipse produced in this way with and without peeling using all times when the Sun was above 30$^\circ$ elevation, which was about 7.5 hours of data.  Before peeling was applied, one can see the impact the bright sources Cas A had on the image; the peeling process seems to have largely mitigated this.

\begin{figure}[h!]
    \centering
         \includegraphics[width = 1.0\linewidth]{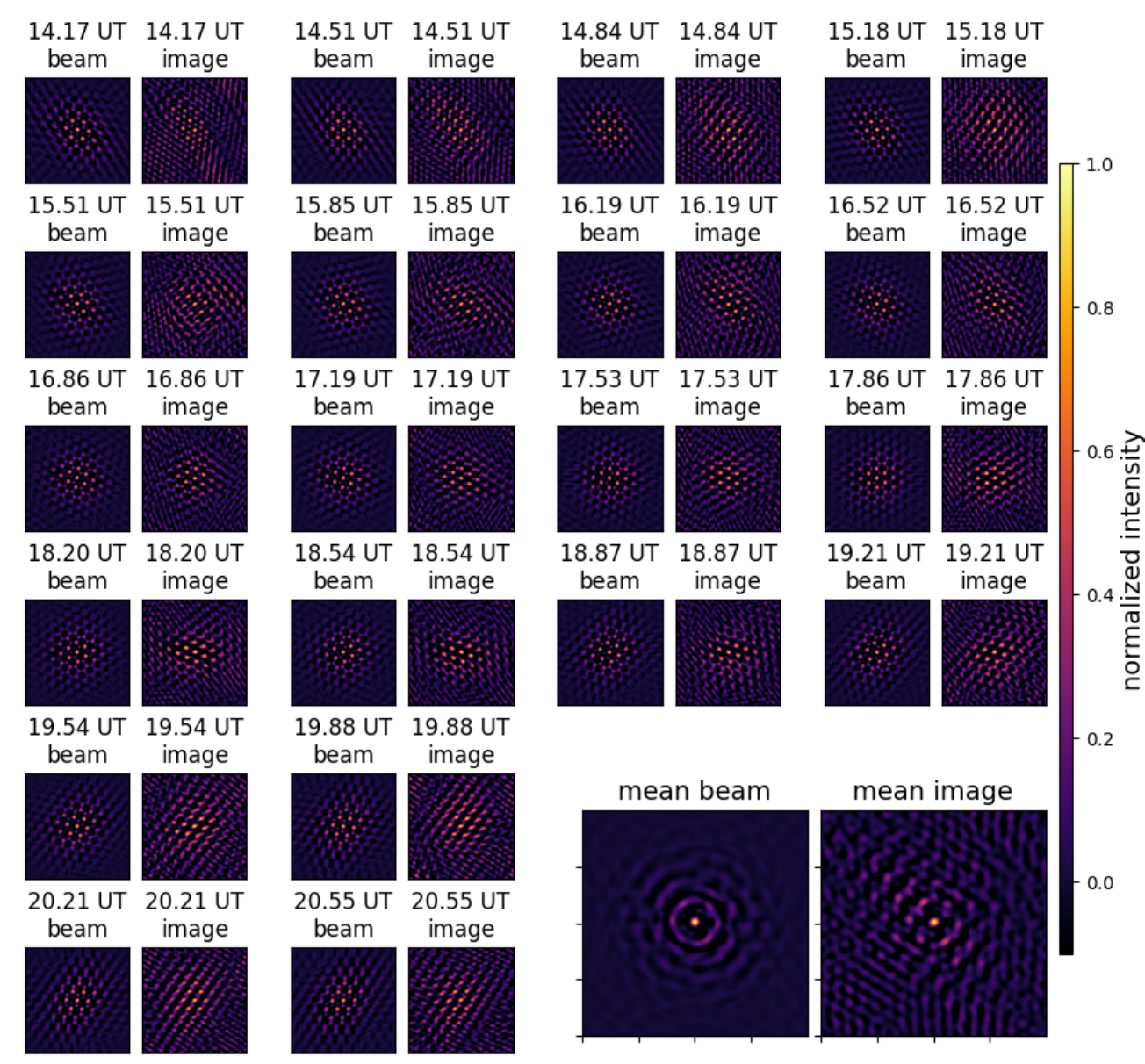}
       \caption{\label{fig:peel_2} Images and theoretical synthesized beams from 20 one-hour segments of data spaced every $\sim$20 minutes during the time the Sun was above $30^\circ$.  Each image is $22.9^\circ$ wide/tall.  The mean image and beam over all those displayed are shown in the lower right panels. }
       
       
  \end{figure}

While the image in the right panel Fig.\ \ref{fig:peel_1} shows that the synthesized beam is rather well-behaved for a four-element array, we note that this is only because the full bandwidth (8.33 MHz) and 7.5 hours of data were used.  For the special case of the eclipse, we are more interested in the temporal variability of the Sun's morphology at these frequencies.  This requires shorter time periods, which will have more poorly behaved beams due to the $u,v$-plane being more sparsely filled by Earth rotation synthesis.  This is illustrated in Fig. \ref{fig:peel_2}, which shows synthesis images every 20 minutes during the same 7.5-hour period, each using one hour of data (i.e., around 40 minutes of overlap among adjacent images).  We found that about one hour of data was needed to reliably and robustly detect the Sun at all times during this period.  Next to each image is the synthesized beam, which includes the known shape of the bandpass as well as the approximate antenna response of $(\sine{e})^{1.6}$ ($e=$elevation).  One can see that with only one hour of data, the $u,v$-coverage is not as good, and the result is many sidelobes near the center.  However, it is also apparent that in all cases, the actual image is quite similar to the theoretical beam, indicating that the Sun was detected and marginally resolved if at all.  The panels in the lower right corner show averages over all 20 images and beams, which show results similar to Fig.\ \ref{fig:peel_1}.

The large number of prominent sidelobes within the one-hour images prohibits the use of an algorithm like CLEAN to reliably remove them.  However, we note that the sidelobe closest to the main lobe are $\sim \! 2.4^\circ$ from the center (i.e., almost 10 solar radii), which is much further out than we would realistically expect to find any bona fide solar emission at the observed frequencies \citep{West_2023}.  Thus, in this context, CLEANing is not necessary, and we may simply examine the central $\sim \! 4.8^\circ$ of each ``dirty'' image.  This is what is shown for each of the 20 one-hour images in the panels of Fig.\ \ref{fig:images}.  For each image, the number of pixels within the main lobe of the calculated beam which were $\ge 1/2$ was used to estimate the full width at half maximum (FWHM), which ranged from $0.63$--$0.67^\circ$ and are indicated in Fig.\ \ref{fig:images} with cyan-colored dashed circles.  Absolute intensity levels for these images were estimated by repeating the imaging process for Cyg A and Cas A (without peeling) and comparing the results with their known intensities (again, see Fig.\ \ref{fig:images}).

\clearpage

\section{RFI Environment Characterization}

\begin{wrapfigure}{r}{0.65\textwidth}
  \begin{center}
    \includegraphics[width=0.65\textwidth]{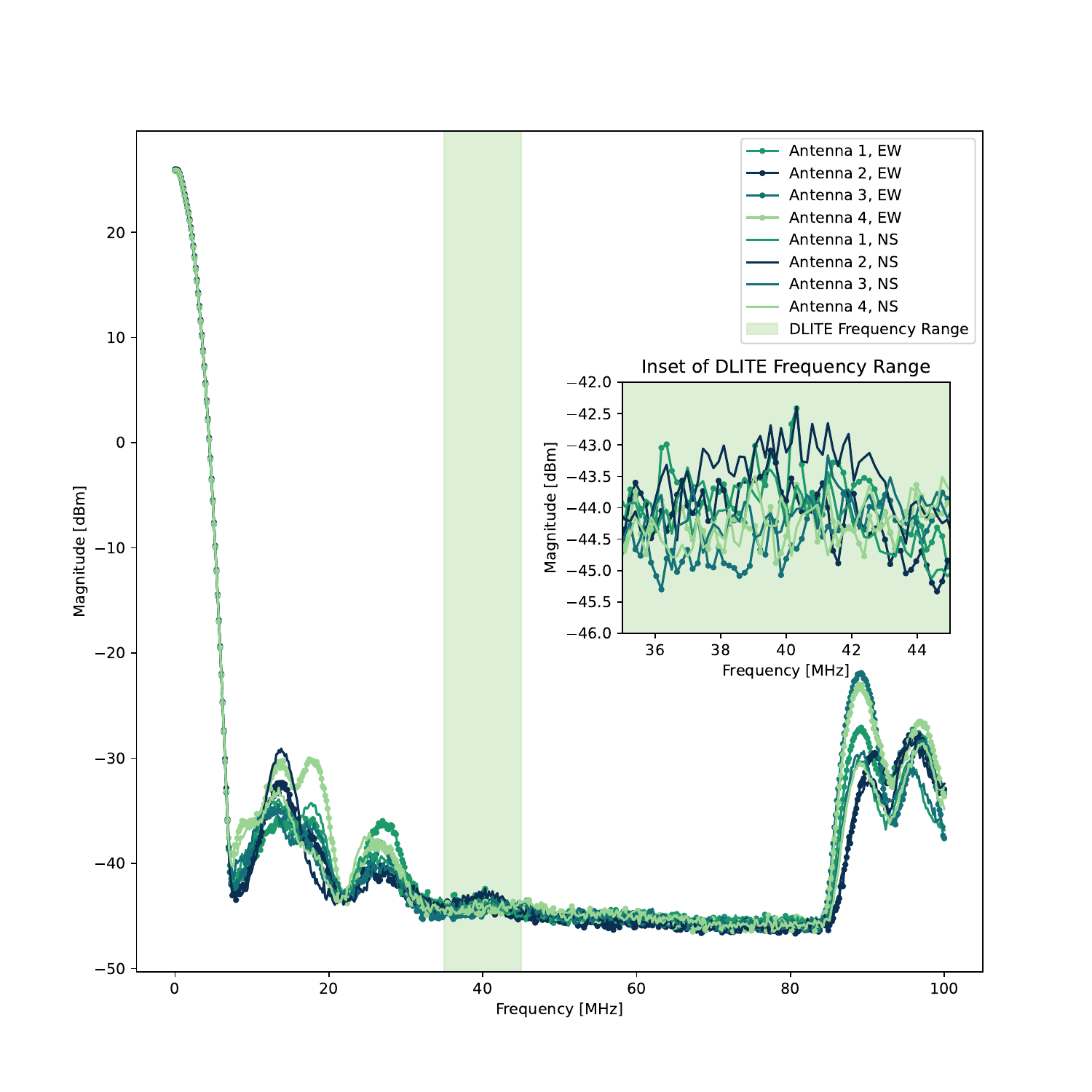}
  \end{center}
  \caption{Site survey RFI measurements taken at Observatory Park on  September 30, 2023 at each of the proposed antenna locations. The frequency range of interest to DLITE in its current configuration is highlighted in the inset plot and indicates a fairly consistent narrow-band noise environment across Observatory Park.}
  \label{fig:RFI-meas}
\end{wrapfigure}

A pilot survey of the RFI environment at Observatory Park was completed on September 30, 2023, to ensure that there were no significant sources of narrow-band RFI at the proposed locations of our four antennas. The methodology outlined in \cite{LWA_memo_RFI} was followed for the measurements. Temperatures were mild ($\sim$75$\degree$ F), with the sky partly cloudy and no significant weather events the entire day. This survey employed an LWA-style antenna and FEE capable of observations from 10 to 120 MHz, a bias tee unit connected to a portable power bank, and a Rohde and Schwarz FSH4-24 spectrum analyzer. Measurements were taken from 0 to 100 MHz with a frequency resolution of 158.7 kHz. For both polarizations in each proposed antenna location, one hundred samples were summed together to create the spectrum shown in Figure \ref{fig:RFI-meas}. These RFI measurements at each of the four proposed antenna locations indicated that there were no prominent narrow-band RFI sources in the 35-45 MHz band at Observatory Park, which would have been seen as prominent spikes in magnitude \citep{LWA_memo_powerlines}. We note that Antenna 2 displayed on average a higher base level of RFI. Antenna 2 is located close to the observatory multi-purpose building, which has a small number of solar panels on the roof, and a transformer box, which is likely the source of the additional RFI (See Figure \ref{fig:DLITE_diagram_maps} for antenna locations). Therefore, the site was deemed as having a relatively quiet narrow-band RFI environment suitable for radio frequency observations. 

This pilot survey was able to determine Observatory Park was free of any significant narrow-band sources of noise that would render one of our baselines unusable for astronomical observations. Observations with the fully commissioned array were still needed to ensure sky noise dominance over the earth-based RFI for each baseline \citep{DLITE}. 

The system equivalent flux density (SEFD) of a single DLITE antenna is a frequency-independent measure of a combination of system noise and collecting area and should be on the order of $2 \times 10^6$ Jy \citep{DLITE, ellingson2013lwa1}. However, as the calibration is approximate and a number of noise sources, such as RFI (e.g., powerlines and lighting) or solar radio bursts can cause drastic fluctuations in SEFDs, a more useful characterization of the noise environment of a site is completed via observations of overall trends over the course of a 24 hour observation.

\begin{wrapfigure}{r}{0.6\textwidth}
  \begin{center}
    \includegraphics[width=0.6\textwidth]{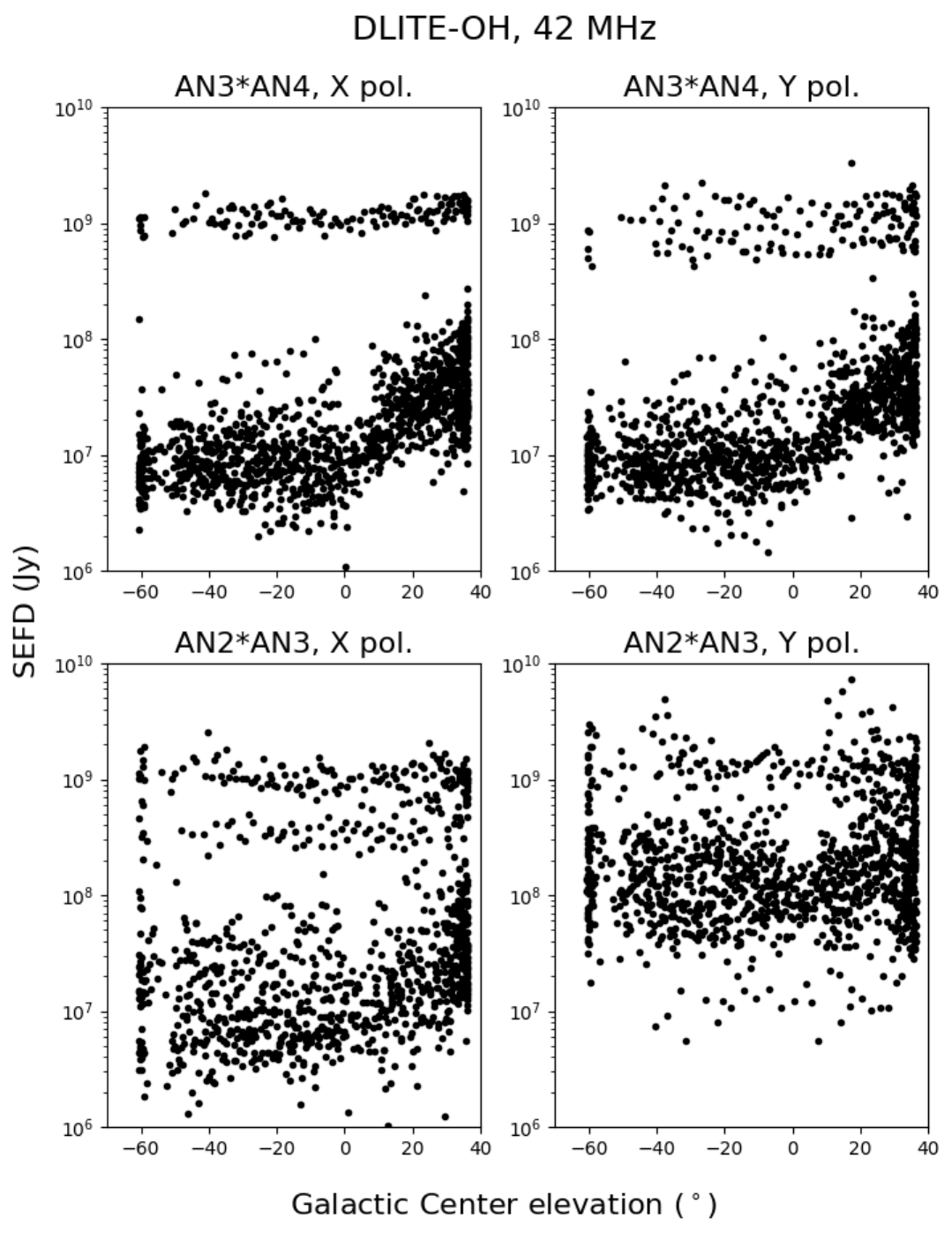}
  \end{center}
  \caption{Trends of noise levels across the course of a 24-hour observation for two of DLITE Ohio's baselines. For both polarizations of the baseline between Antenna 3 and 4 and the NS (X) polarization baseline between Antenna 2 and 3 to a lesser extent, as the galactic center rises overhead, the noise floor sweeps upward, indicating that these baselines are sky-noise-dominated. These baselines still experience individual noise features, as indicated by the thin band across the top of the plots. These system equivalent flux density (SEFD) features can be caused by a number of short-term RFI events, such as lighting or solar bursts \citep{DLITE}. We do not see this trend overall for the EW (Y) polarization of the baseline between Antenna 2 and 3, which is likely due to the elevated noise levels in the EW polarization at Antenna 2. Therefore, this baseline is not sky-noise-dominated.  }
  \label{fig:RFI-sefd}
\end{wrapfigure}

Galactic noise is the largest source of sky noise at these frequencies, meaning that if DLITE Ohio observations are sufficiently dominated by sky noise the average noise should in general follow an increasing trend as the galaxy rises overhead (see section 3.3 and Figure 17 in \cite{DLITE}). This trend can be seen over the course of a full day of observations, as exhibited by selected baselines in Figure \ref{fig:RFI-sefd}. 

In both the NS (X) and EW (Y) polarizations of the baseline between Antenna 3 and 4, we see our average RFI increase as the galaxy rises over DLITE Ohio, indicating that these baselines are sky-noise-dominated. We see a similar trend but with significantly less dominance in the NS  polarization of our baseline between Antenna 2 and 3. However, in the EW polarization four baselines between Antenna 2 and 3, we see no such trend, meaning that during the duration of this RFI environment survey, this baseline was earth-noise-dominated. Referring back to Figure \ref{fig:RFI-meas}, we note that the EW polarization of Antenna 2 experienced on average greater levels of RFI, which corroborates the findings of this larger survey.

\clearpage

\section{Parts List}

This appendix contains a complete list of all hardware components used to construct DLITE Ohio, along with their manufacturer and current prices from the time of purchase in Spring 2024. The total price of these components was $\sim$\$42,800.00 US, not including shipping or sales tax. A significant fraction of this cost was from the large amounts of LMR-400 coaxial cable ($\sim$\$11,000 US) and the four Software Defined Radios ($\sim$\$14,000 US) required. Therefore, there are efforts to upgrade the DLITE system to operate on less expensive but comparable components.

\begin{longtblr}[
  caption = {Parts Used to construct DLITE Ohio},
  label = {tab:parts},
]{
  colspec = {|X|X|X|X|X|},
  rowhead = 1,
  hlines,
} 
\bf{Item:} & \bf{Number Used:} & \bf{Manufacturer:} & \bf{Current price:} & \bf{Notes:} \\

Inverted vee dipole antennas & 4 & Burns Industries & US $\$495.00 $ per antenna & Antennas used for LWA, includes mast\\
Front end electronics (FEEs)  & 4 & Burns Industries & US $\$230.00$ per unit \\

Ground stake & 4 & Burns Industries & US $\$41.00$ per unit & All antenna elements (dipole antenna, FEE, and ground stake) can also be purchased as a kit from Reeve Radio Observatories \\ 

Times Microwave Systems Inc. LMR-240DB Flex Cable & 8 & TESSCO Technologies Incorporated & US $\$1.07$ per foot & Used to connect FEEs to longer baselines, $\sim$2 meters per polarization \\

Times Microwave Systems Inc. SMA Male for LMR-240DB Flex Cable  & 8 & TESSCO Technologies Incorporated & US $\$8.47$ per unit\\

Times Microwave Systems Inc. N Male for LMR-240DB Flex Cable  &  8 & TESSCO Technologies Incorporated & US $\$10.79$ per unit \\

Times Microwave Systems Inc. N Female Hex/Knurl Combo No braid trim for LMR-400 Cable  & 16 & TESSCO Technologies Incorporated & US $\$12.36$ per unit \\

Times Microwave Systems Inc. 3/8" LMR-400 Flooded & 7,500 & TESSCO Technologies Incorporated & US $\$1.45$ per foot & Divided into 8 baselines of different lengths, see section \\

8 channel bias tee & 1 & Galaxy Electronics & US $\$482.00$ per unit & Part number: $Bias\_T\_8\_06$ \\

Software Defined Radio, USRP N210 & 4 & Ettus Research & US $\$3,455.00$ per unit \\

BasicRX 1-250 MHz Daughterboard  & 4 & Ettus Research & US $\$154.00$ per unit\\ 
OctoClock-G CDA-2990 & 1  & Ettus Research & US $\$3,492.00$ per unit \\

Low Pass Filter  & 8  & Mini Circuits  & US $\$48.64$ per unit \\

SMA (LMR-400 to bias tee) & 8 & Mini Circuits & US $\$178.44$ per unit & 141-10SMNB+ Hand-Flex Interconnect, 0.141" center diameter, 12.5 GHz \\

Junction Boxes & 4 & Burns Industries & US $\$12.00$ per unit \\

SMA (bias tee to low pass filters/SDR, from SDR to clock)  & 16 & Mini-circuits  & US $\$19.57$ per unit & 086-10SM+
Hand-Flex Interconnect, 0.086" center diameter, 18.0 GHz \\

Ethernet Cable & 4 & Tessco & US $\$18.67$ per unit & C2G 6ft Cat6 Ethernet Cable \\

Pelican case & 1 & Rack Case Solutions  &US $\$2,190.89$
per unit & Holds full backend end and computer \\

DC Power Supply   & 1 & Acopian  & US $\$1,670$ per unit & Model Y0100LXU720C1E1M3\-DIO1 \\

Rack mounted computer  & 1 & SuperLogics &US $\$2,494$ per unit & Model SL-2U-MH510I-GD\\

Moniter/keyboard drawer & 1 & SuperLogics  & US $\$924$ per unit & Model SL-RMPD-RP-119 \\

Rack-mounted powerstrip & 1 & Digi-Key & US $\$106.00$ per unit & Tripp Lite model RS1215-RA \\

Lightning Arrestor & 8 & Tessco & US $\$43.78$ per unit & 0-7 GHz Lighting Arrestor with N Female - N Female Bulkhead connectors \\

\end{longtblr}

\end{document}

%% file: DLITE_system_diagram
\tikzset{every picture/.style={line width=0.75pt}} 

\begin{tikzpicture}[x=0.75pt,y=0.75pt,yscale=-1,xscale=1]
\draw [line width=2.25]    (39.7,41.19) -- (40.37,100.3) ;
\draw [line width=1.5]    (39.7,41.19) -- (10.37,80.07) ;
\draw [line width=1.5]    (39.7,41.19) -- (69.93,80.3) ;
\draw    (39.7,41.19) -- (24.59,80.07) ;
\draw    (39.7,41.19) -- (55.26,79.85) ;
\draw    (24.59,80.07) -- (55.26,79.85) ;
\draw    (30.59,64.07) -- (49.26,64.07) ;
\draw  [color={rgb, 255:red, 13; green, 98; blue, 199 }  ,draw opacity=1 ][fill={rgb, 255:red, 12; green, 90; blue, 182 }  ,fill opacity=1 ] (30,30.54) -- (49.7,30.54) -- (49.7,40.96) -- (30,40.96) -- cycle ;
\draw [line width=2.25]    (39.95,120.69) -- (40.62,179.8) ;
\draw [line width=1.5]    (39.95,120.69) -- (10.62,159.57) ;
\draw [line width=1.5]    (39.95,120.69) -- (70.18,159.8) ;
\draw    (39.95,120.69) -- (24.84,159.57) ;
\draw    (39.95,120.69) -- (55.51,159.35) ;
\draw    (24.84,159.57) -- (55.51,159.35) ;
\draw    (30.84,143.57) -- (49.51,143.57) ;
\draw  [color={rgb, 255:red, 13; green, 98; blue, 199 }  ,draw opacity=1 ][fill={rgb, 255:red, 12; green, 90; blue, 182 }  ,fill opacity=1 ] (30.25,110.04) -- (49.95,110.04) -- (49.95,120.46) -- (30.25,120.46) -- cycle ;
\draw [line width=2.25]    (39.95,201.19) -- (40.62,260.3) ;
\draw [line width=1.5]    (39.95,201.19) -- (10.62,240.07) ;
\draw [line width=1.5]    (39.95,201.19) -- (70.18,240.3) ;
\draw    (39.95,201.19) -- (24.84,240.07) ;
\draw    (39.95,201.19) -- (55.51,239.85) ;
\draw    (24.84,240.07) -- (55.51,239.85) ;
\draw    (30.84,224.07) -- (49.51,224.07) ;
\draw  [color={rgb, 255:red, 13; green, 98; blue, 199 }  ,draw opacity=1 ][fill={rgb, 255:red, 12; green, 90; blue, 182 }  ,fill opacity=1 ] (30.25,190.54) -- (49.95,190.54) -- (49.95,200.96) -- (30.25,200.96) -- cycle ;
\draw [line width=2.25]    (40.2,281.44) -- (40.87,340.55) ;
\draw [line width=1.5]    (40.2,281.44) -- (10.87,320.32) ;
\draw [line width=1.5]    (40.2,281.44) -- (70.43,320.55) ;
\draw    (40.2,281.44) -- (25.09,320.32) ;
\draw    (40.2,281.44) -- (55.76,320.1) ;
\draw    (25.09,320.32) -- (55.76,320.1) ;
\draw    (31.09,304.32) -- (49.76,304.32) ;
\draw  [color={rgb, 255:red, 13; green, 98; blue, 199 }  ,draw opacity=1 ][fill={rgb, 255:red, 12; green, 90; blue, 182 }  ,fill opacity=1 ] (30.5,270.79) -- (50.2,270.79) -- (50.2,281.21) -- (30.5,281.21) -- cycle ;
\draw  [color={rgb, 255:red, 155; green, 243; blue, 57 }  ,draw opacity=1 ][fill={rgb, 255:red, 65; green, 117; blue, 5 }  ,fill opacity=1 ] (250.69,59.94) -- (249.95,361.27) -- (209.95,361.18) -- (210.69,59.84) -- cycle ;
\draw [line width=1.5]    (40.37,100.3) -- (83.11,79.67) -- (207.78,81.17) ;
\draw [shift={(211.78,81.22)}, rotate = 180.69] [fill={rgb, 255:red, 0; green, 0; blue, 0 }  ][line width=0.08]  [draw opacity=0] (11.61,-5.58) -- (0,0) -- (11.61,5.58) -- cycle    ;
\draw [line width=1.5]    (40.37,100.3) -- (206.44,100.98) ;
\draw [shift={(210.44,101)}, rotate = 180.24] [fill={rgb, 255:red, 0; green, 0; blue, 0 }  ][line width=0.08]  [draw opacity=0] (11.61,-5.58) -- (0,0) -- (11.61,5.58) -- cycle    ;
\draw [line width=1.5]    (40.62,179.8) -- (83.78,161.22) -- (205.78,160.58) ;
\draw [shift={(209.78,160.56)}, rotate = 179.7] [fill={rgb, 255:red, 0; green, 0; blue, 0 }  ][line width=0.08]  [draw opacity=0] (11.61,-5.58) -- (0,0) -- (11.61,5.58) -- cycle    ;
\draw [line width=1.5]    (40.62,179.8) -- (206.44,181.19) ;
\draw [shift={(210.44,181.22)}, rotate = 180.48] [fill={rgb, 255:red, 0; green, 0; blue, 0 }  ][line width=0.08]  [draw opacity=0] (11.61,-5.58) -- (0,0) -- (11.61,5.58) -- cycle    ;
\draw [line width=1.5]    (40.62,260.3) -- (206.44,260.33) ;
\draw [shift={(210.44,260.33)}, rotate = 180.01] [fill={rgb, 255:red, 0; green, 0; blue, 0 }  ][line width=0.08]  [draw opacity=0] (11.61,-5.58) -- (0,0) -- (11.61,5.58) -- cycle    ;
\draw [line width=1.5]    (40.62,260.3) -- (83.11,240.56) -- (206.44,240.56) ;
\draw [shift={(210.44,240.56)}, rotate = 180] [fill={rgb, 255:red, 0; green, 0; blue, 0 }  ][line width=0.08]  [draw opacity=0] (11.61,-5.58) -- (0,0) -- (11.61,5.58) -- cycle    ;
\draw [line width=1.5]    (40.87,340.55) -- (83.78,320.78) -- (205.78,320.78) ;
\draw [shift={(209.78,320.78)}, rotate = 180] [fill={rgb, 255:red, 0; green, 0; blue, 0 }  ][line width=0.08]  [draw opacity=0] (11.61,-5.58) -- (0,0) -- (11.61,5.58) -- cycle    ;
\draw [line width=1.5]    (40.87,340.55) -- (206.44,340.56) ;
\draw [shift={(210.44,340.56)}, rotate = 180] [fill={rgb, 255:red, 0; green, 0; blue, 0 }  ][line width=0.08]  [draw opacity=0] (11.61,-5.58) -- (0,0) -- (11.61,5.58) -- cycle    ;
\draw [color={rgb, 255:red, 189; green, 16; blue, 224 }  ,draw opacity=1 ][line width=0.75]    (258.33,79.5) -- (437.5,80.08) ;
\draw [shift={(259.33,79.51)}, rotate = 180.19] [fill={rgb, 255:red, 189; green, 16; blue, 224 }  ,fill opacity=1 ][line width=0.08]  [draw opacity=0] (8.93,-4.29) -- (0,0) -- (8.93,4.29) -- cycle    ;
\draw  [color={rgb, 255:red, 245; green, 166; blue, 35 }  ,draw opacity=1 ][fill={rgb, 255:red, 248; green, 231; blue, 28 }  ,fill opacity=1 ] (320.93,71.08) -- (339.07,71.08) .. controls (341.42,71.08) and (343.33,75.11) .. (343.33,80.08) .. controls (343.33,85.05) and (341.42,89.08) .. (339.07,89.08) -- (320.93,89.08) .. controls (318.58,89.08) and (316.67,85.05) .. (316.67,80.08) .. controls (316.67,75.11) and (318.58,71.08) .. (320.93,71.08) -- cycle ;
\draw [color={rgb, 255:red, 189; green, 16; blue, 224 }  ,draw opacity=1 ][line width=0.75]    (258.33,99.5) -- (438,100.08) ;
\draw [shift={(259.33,99.51)}, rotate = 180.18] [fill={rgb, 255:red, 189; green, 16; blue, 224 }  ,fill opacity=1 ][line width=0.08]  [draw opacity=0] (8.93,-4.29) -- (0,0) -- (8.93,4.29) -- cycle    ;
\draw  [color={rgb, 255:red, 245; green, 166; blue, 35 }  ,draw opacity=1 ][fill={rgb, 255:red, 248; green, 231; blue, 28 }  ,fill opacity=1 ] (320.73,90.88) -- (338.87,90.88) .. controls (341.22,90.88) and (343.13,94.91) .. (343.13,99.88) .. controls (343.13,104.85) and (341.22,108.88) .. (338.87,108.88) -- (320.73,108.88) .. controls (318.38,108.88) and (316.47,104.85) .. (316.47,99.88) .. controls (316.47,94.91) and (318.38,90.88) .. (320.73,90.88) -- cycle ;
\draw [color={rgb, 255:red, 189; green, 16; blue, 224 }  ,draw opacity=1 ][line width=0.75]    (258.33,160.49) -- (437.17,159.33) ;
\draw [shift={(259.33,160.49)}, rotate = 179.63] [fill={rgb, 255:red, 189; green, 16; blue, 224 }  ,fill opacity=1 ][line width=0.08]  [draw opacity=0] (8.93,-4.29) -- (0,0) -- (8.93,4.29) -- cycle    ;
\draw  [color={rgb, 255:red, 245; green, 166; blue, 35 }  ,draw opacity=1 ][fill={rgb, 255:red, 248; green, 231; blue, 28 }  ,fill opacity=1 ] (321.93,150.58) -- (340.07,150.58) .. controls (342.42,150.58) and (344.33,154.61) .. (344.33,159.58) .. controls (344.33,164.55) and (342.42,168.58) .. (340.07,168.58) -- (321.93,168.58) .. controls (319.58,168.58) and (317.67,164.55) .. (317.67,159.58) .. controls (317.67,154.61) and (319.58,150.58) .. (321.93,150.58) -- cycle ;
\draw [color={rgb, 255:red, 189; green, 16; blue, 224 }  ,draw opacity=1 ][line width=0.75]    (258.33,179.5) -- (436.67,178.83) ;
\draw [shift={(259.33,179.49)}, rotate = 179.79] [fill={rgb, 255:red, 189; green, 16; blue, 224 }  ,fill opacity=1 ][line width=0.08]  [draw opacity=0] (8.93,-4.29) -- (0,0) -- (8.93,4.29) -- cycle    ;
\draw  [color={rgb, 255:red, 245; green, 166; blue, 35 }  ,draw opacity=1 ][fill={rgb, 255:red, 248; green, 231; blue, 28 }  ,fill opacity=1 ] (321.43,170.08) -- (339.57,170.08) .. controls (341.92,170.08) and (343.83,174.11) .. (343.83,179.08) .. controls (343.83,184.05) and (341.92,188.08) .. (339.57,188.08) -- (321.43,188.08) .. controls (319.08,188.08) and (317.17,184.05) .. (317.17,179.08) .. controls (317.17,174.11) and (319.08,170.08) .. (321.43,170.08) -- cycle ;
\draw [color={rgb, 255:red, 189; green, 16; blue, 224 }  ,draw opacity=1 ][line width=0.75]    (258.33,239.4) -- (436.17,240.08) ;
\draw [shift={(259.33,239.41)}, rotate = 180.22] [fill={rgb, 255:red, 189; green, 16; blue, 224 }  ,fill opacity=1 ][line width=0.08]  [draw opacity=0] (8.93,-4.29) -- (0,0) -- (8.93,4.29) -- cycle    ;
\draw  [color={rgb, 255:red, 245; green, 166; blue, 35 }  ,draw opacity=1 ][fill={rgb, 255:red, 248; green, 231; blue, 28 }  ,fill opacity=1 ] (320.93,231.33) -- (339.07,231.33) .. controls (341.42,231.33) and (343.33,235.36) .. (343.33,240.33) .. controls (343.33,245.3) and (341.42,249.33) .. (339.07,249.33) -- (320.93,249.33) .. controls (318.58,249.33) and (316.67,245.3) .. (316.67,240.33) .. controls (316.67,235.36) and (318.58,231.33) .. (320.93,231.33) -- cycle ;
\draw [color={rgb, 255:red, 189; green, 16; blue, 224 }  ,draw opacity=1 ][line width=0.75]    (258.33,260.2) -- (436.5,259.42) ;
\draw [shift={(259.33,260.19)}, rotate = 179.75] [fill={rgb, 255:red, 189; green, 16; blue, 224 }  ,fill opacity=1 ][line width=0.08]  [draw opacity=0] (8.93,-4.29) -- (0,0) -- (8.93,4.29) -- cycle    ;
\draw  [color={rgb, 255:red, 245; green, 166; blue, 35 }  ,draw opacity=1 ][fill={rgb, 255:red, 248; green, 231; blue, 28 }  ,fill opacity=1 ] (320.43,250.83) -- (338.57,250.83) .. controls (340.92,250.83) and (342.83,254.86) .. (342.83,259.83) .. controls (342.83,264.8) and (340.92,268.83) .. (338.57,268.83) -- (320.43,268.83) .. controls (318.08,268.83) and (316.17,264.8) .. (316.17,259.83) .. controls (316.17,254.86) and (318.08,250.83) .. (320.43,250.83) -- cycle ;
\draw [color={rgb, 255:red, 189; green, 16; blue, 224 }  ,draw opacity=1 ][line width=0.75]    (257.93,320.6) -- (438,320.08) ;
\draw [shift={(258.93,320.59)}, rotate = 179.84] [fill={rgb, 255:red, 189; green, 16; blue, 224 }  ,fill opacity=1 ][line width=0.08]  [draw opacity=0] (8.93,-4.29) -- (0,0) -- (8.93,4.29) -- cycle    ;
\draw  [color={rgb, 255:red, 245; green, 166; blue, 35 }  ,draw opacity=1 ][fill={rgb, 255:red, 248; green, 231; blue, 28 }  ,fill opacity=1 ] (320.93,311.33) -- (339.07,311.33) .. controls (341.42,311.33) and (343.33,315.36) .. (343.33,320.33) .. controls (343.33,325.3) and (341.42,329.33) .. (339.07,329.33) -- (320.93,329.33) .. controls (318.58,329.33) and (316.67,325.3) .. (316.67,320.33) .. controls (316.67,315.36) and (318.58,311.33) .. (320.93,311.33) -- cycle ;
\draw [color={rgb, 255:red, 189; green, 16; blue, 224 }  ,draw opacity=1 ][line width=0.75]    (257.93,340.46) -- (438,340.08) ;
\draw [shift={(258.93,340.46)}, rotate = 179.88] [fill={rgb, 255:red, 189; green, 16; blue, 224 }  ,fill opacity=1 ][line width=0.08]  [draw opacity=0] (8.93,-4.29) -- (0,0) -- (8.93,4.29) -- cycle    ;
\draw  [color={rgb, 255:red, 245; green, 166; blue, 35 }  ,draw opacity=1 ][fill={rgb, 255:red, 248; green, 231; blue, 28 }  ,fill opacity=1 ] (320.43,330.83) -- (338.57,330.83) .. controls (340.92,330.83) and (342.83,334.86) .. (342.83,339.83) .. controls (342.83,344.8) and (340.92,348.83) .. (338.57,348.83) -- (320.43,348.83) .. controls (318.08,348.83) and (316.17,344.8) .. (316.17,339.83) .. controls (316.17,334.86) and (318.08,330.83) .. (320.43,330.83) -- cycle ;
\draw  [color={rgb, 255:red, 74; green, 74; blue, 74 }  ,draw opacity=1 ][fill={rgb, 255:red, 155; green, 155; blue, 155 }  ,fill opacity=1 ] (438,59.75) -- (498,59.75) -- (498,120.75) -- (438,120.75) -- cycle ; \draw  [color={rgb, 255:red, 74; green, 74; blue, 74 }  ,draw opacity=1 ] (445.5,59.75) -- (445.5,120.75) ; \draw  [color={rgb, 255:red, 74; green, 74; blue, 74 }  ,draw opacity=1 ] (490.5,59.75) -- (490.5,120.75) ;
\draw  [color={rgb, 255:red, 74; green, 74; blue, 74 }  ,draw opacity=1 ][fill={rgb, 255:red, 155; green, 155; blue, 155 }  ,fill opacity=1 ] (437,139.25) -- (497,139.25) -- (497,200.25) -- (437,200.25) -- cycle ; \draw  [color={rgb, 255:red, 74; green, 74; blue, 74 }  ,draw opacity=1 ] (444.5,139.25) -- (444.5,200.25) ; \draw  [color={rgb, 255:red, 74; green, 74; blue, 74 }  ,draw opacity=1 ] (489.5,139.25) -- (489.5,200.25) ;
\draw  [color={rgb, 255:red, 74; green, 74; blue, 74 }  ,draw opacity=1 ][fill={rgb, 255:red, 155; green, 155; blue, 155 }  ,fill opacity=1 ] (437,220.25) -- (497,220.25) -- (497,281.25) -- (437,281.25) -- cycle ; \draw  [color={rgb, 255:red, 74; green, 74; blue, 74 }  ,draw opacity=1 ] (444.5,220.25) -- (444.5,281.25) ; \draw  [color={rgb, 255:red, 74; green, 74; blue, 74 }  ,draw opacity=1 ] (489.5,220.25) -- (489.5,281.25) ;
\draw  [color={rgb, 255:red, 74; green, 74; blue, 74 }  ,draw opacity=1 ][fill={rgb, 255:red, 155; green, 155; blue, 155 }  ,fill opacity=1 ] (438,300.25) -- (498,300.25) -- (498,361.25) -- (438,361.25) -- cycle ; \draw  [color={rgb, 255:red, 74; green, 74; blue, 74 }  ,draw opacity=1 ] (445.5,300.25) -- (445.5,361.25) ; \draw  [color={rgb, 255:red, 74; green, 74; blue, 74 }  ,draw opacity=1 ] (490.5,300.25) -- (490.5,361.25) ;
\draw  [color={rgb, 255:red, 0; green, 0; blue, 0 }  ,draw opacity=1 ][fill={rgb, 255:red, 155; green, 155; blue, 155 }  ,fill opacity=0.52 ][line width=1.5]  (621,138.33) -- (681,138.33) -- (681,272.83) -- (621,272.83) -- cycle ; \draw  [color={rgb, 255:red, 0; green, 0; blue, 0 }  ,draw opacity=1 ][line width=1.5]  (629.33,146.67) -- (672.67,146.67) -- (672.67,264.5) -- (629.33,264.5) -- cycle ; \draw  [color={rgb, 255:red, 0; green, 0; blue, 0 }  ,draw opacity=1 ][line width=1.5]  (621,138.33) -- (629.33,146.67) ; \draw  [color={rgb, 255:red, 0; green, 0; blue, 0 }  ,draw opacity=1 ][line width=1.5]  (681,138.33) -- (672.67,146.67) ; \draw  [color={rgb, 255:red, 0; green, 0; blue, 0 }  ,draw opacity=1 ][line width=1.5]  (681,272.83) -- (672.67,264.5) ; \draw  [color={rgb, 255:red, 0; green, 0; blue, 0 }  ,draw opacity=1 ][line width=1.5]  (621,272.83) -- (629.33,264.5) ;
\draw [color={rgb, 255:red, 74; green, 144; blue, 226 }  ,draw opacity=1 ][line width=2.25]    (539.9,41.24) -- (570.19,41.24) ;
\draw [line width=1.5]    (539.71,25.14) -- (570.19,25.24) ;
\draw  [color={rgb, 255:red, 155; green, 155; blue, 155 }  ,draw opacity=1 ][fill={rgb, 255:red, 128; green, 128; blue, 128 }  ,fill opacity=0.69 ] (626.83,422.33) .. controls (623.33,422.33) and (620.49,419.49) .. (620.49,415.99) -- (620.51,306.67) .. controls (620.51,303.17) and (623.35,300.33) .. (626.85,300.33) -- (645.89,300.33) .. controls (649.39,300.33) and (652.23,303.18) .. (652.23,306.68) -- (652.21,415.99) .. controls (652.21,419.49) and (649.37,422.33) .. (645.87,422.33) -- cycle ;
\draw [color={rgb, 255:red, 189; green, 16; blue, 224 }  ,draw opacity=1 ]   (498.48,120.17) -- (570,120.42) -- (570.5,320.42) -- (619.86,320.31) ;
\draw [color={rgb, 255:red, 189; green, 16; blue, 224 }  ,draw opacity=1 ]   (497.9,200.21) -- (550.19,200.21) -- (549.9,349.88) -- (619.44,350.44) ;
\draw [color={rgb, 255:red, 189; green, 16; blue, 224 }  ,draw opacity=1 ]   (497.62,190.21) -- (559.47,190.6) -- (559.76,340.31) -- (619.9,340.45) ;
\draw [color={rgb, 255:red, 189; green, 16; blue, 224 }  ,draw opacity=1 ]   (497.76,280.4) -- (530.05,280.4) -- (529.47,380.2) -- (619.47,380.33) ;
\draw [color={rgb, 255:red, 189; green, 16; blue, 224 }  ,draw opacity=1 ]   (498.62,110.6) -- (580,110.42) -- (580,309.92) -- (620,310.45) ;
\draw [color={rgb, 255:red, 189; green, 16; blue, 224 }  ,draw opacity=1 ]   (497.62,270.26) -- (540.19,270.26) -- (539.87,369.8) -- (619.87,370.33) ;
\draw [color={rgb, 255:red, 189; green, 16; blue, 224 }  ,draw opacity=1 ]   (498.48,360.21) -- (510.19,360.21) -- (510.27,410.6) -- (619.87,410.6) ;
\draw [color={rgb, 255:red, 189; green, 16; blue, 224 }  ,draw opacity=1 ]   (498.8,350.45) -- (520.67,350.6) -- (520.19,400.21) -- (620.67,400.6) ;
\draw [color={rgb, 255:red, 189; green, 16; blue, 224 }  ,draw opacity=1 ]   (540.97,55.17) -- (571.22,55.21) ;
\draw [color={rgb, 255:red, 74; green, 144; blue, 226 }  ,draw opacity=1 ][line width=2.25]    (499.33,89.75) -- (551.67,89.25) -- (551.67,170.25) -- (621,170.67) ;
\draw [color={rgb, 255:red, 74; green, 144; blue, 226 }  ,draw opacity=1 ][line width=2.25]    (498,170.33) -- (541.05,170.93) -- (540.9,189.79) -- (620.9,189.93) ;
\draw [color={rgb, 255:red, 74; green, 144; blue, 226 }  ,draw opacity=1 ][line width=2.25]    (497.62,249.98) -- (541.33,249.83) -- (541.33,229.83) -- (620.17,229.75) ;
\draw [color={rgb, 255:red, 74; green, 144; blue, 226 }  ,draw opacity=1 ][line width=2.25]    (499.33,332.83) -- (551.33,332.83) -- (551.33,251.83) -- (620.17,251.75) ;
\draw (222.64,268.28) node [anchor=north west][inner sep=0.75pt]  [color={rgb, 255:red, 255; green, 255; blue, 255 }  ,opacity=1 ,rotate=-270.35] [align=left] {8-Channel Bias-T};
\draw (285,45.33) node [anchor=north west][inner sep=0.75pt]  [font=\large] [align=left] {{\footnotesize Low Pass Filters} };
\draw (446,34.33) node [anchor=north west][inner sep=0.75pt]   [align=left] {SDRs};
\draw (640.88,241.81) node [anchor=north west][inner sep=0.75pt]  [font=\large,rotate=-270.2] [align=left] {Computer };
\draw (580,18) node [anchor=north west][inner sep=0.75pt]   [align=left] {Coaxial Cable \\Ethernet Cable\\SMA Cable};
\draw (631.82,404.83) node [anchor=north west][inner sep=0.75pt]  [font=\footnotesize,color={rgb, 255:red, 255; green, 255; blue, 255 }  ,opacity=1 ,rotate=-270.01] [align=left] {Reference Clock};
\end{tikzpicture}